\documentclass[journal,12pt,onecolumn,final]{IEEEtran}
\pdfoutput=1
\usepackage{times}
\usepackage{cite}
\usepackage{mdwmath}
\usepackage[english]{babel}
\usepackage{epsfig}
\usepackage{easymat}
\usepackage{amsfonts}
\usepackage{bm}
\usepackage{amsbsy}
\usepackage{amsmath}
\usepackage{float}
\usepackage{graphicx}
\usepackage{balance}
\usepackage{latexsym}
\usepackage{amssymb}
\usepackage{txfonts}
\usepackage{wasysym}
\usepackage{mathbbol}
\usepackage{multirow}
\usepackage{stfloats}
\usepackage{enumerate}
\usepackage{flushend}   
\usepackage{amsfonts}
\usepackage{setspace}
\usepackage{graphicx}
\usepackage{caption}
\usepackage{subcaption}
\usepackage[procnumbered,linesnumbered,ruled,vlined]{algorithm2e}
\usepackage{amsmath}
\usepackage{tikz} 
\usetikzlibrary{patterns}
\usepackage{pgfplots}

\usepackage{comment}
\def\baselinestretch{0.92}

\title{Performance / Complexity Trade-offs of the Sphere Decoder Algorithm for Massive MIMO Systems}

\author{\IEEEauthorblockN{A. Dabah\IEEEauthorrefmark{1},
H. Ltaief\IEEEauthorrefmark{1},
Z. Rezki\IEEEauthorrefmark{2},
M.-A. Arfaoui\IEEEauthorrefmark{3},
M.-S. Alouini\IEEEauthorrefmark{1}, and
D. Keyes\IEEEauthorrefmark{1}}
\\~
\\~
\IEEEauthorblockA{\IEEEauthorrefmark{1}Computer, Electrical and Mathematical Science and Engineering, \\King Abdullah University of Science and Technology\\
\{Adel.Dabah.1, Hatem.Ltaief, Slim.Alouini, David.Keyes\}@kaust.edu.sa}; Adabah@cerist.dz
\\~
\\~
\IEEEauthorblockA{\IEEEauthorrefmark{2}University of Idaho, Moscow, ID USA,\\
zrezki@uidaho.edu}
\\~
\\~
\IEEEauthorblockA{\IEEEauthorrefmark{3}Concordia University, Montreal, Canada,\\
m\_arfaou@encs.concordia.ca}}

\begin{document}
\maketitle

\begin{abstract} 
Massive Multiple-Input Multiple-Output (Massive MIMO) systems are seen by many researchers 
as a paramount technology toward next generation networks. This technology consists 
of hundreds of antennas that are capable of sending and receiving simultaneously a huge amount of data. 
One of the main challenges when using this technology is the necessity  
of an efficient decoding framework. The latter must guarantee both a real-time response 
complexity and a good Bit Error Rate (BER) performance.
The Sphere Decoder (SD) algorithm represents one of the promising Maximum Likelihood (ML)
decoding algorithms in terms of BER. 
However, it is inefficient for dealing with large MIMO systems due to its prohibitive complexity.
To overcome this drawback, we propose to revisit the sequential SD algorithm
and implement several variants that aim 
at finding appropriate trade-offs between complexity and performance. 
We conduct experiments to assess the critical impact of the SD
components, i.e., the exploration strategies and the evaluation process, and to
accelerate the search for the optimal combination of the transmitted vector. 
Then, we propose an efficient high-level parallel SD scheme based on the master/worker paradigm,
which permits multiple SD instances to simultaneously explore the search tree, while
mitigating the overheads from load imbalance. 
The results of our parallel SD implementation outperform the state-of-the-art
by more than 5$\times$ using similar MIMO configuration systems, and show a super-linear speedup
on multicore platforms.
Moreover, this paper presents a new hybrid
implementation, which enables to simulate MIMO configurations at an even larger scale.
It combines the strengths of SD and K-best algorithms, i.e., by maintaining the low BER of SD, 
while further reducing the complexity using the K-best way of pruning search space.
The hybrid approach extends our parallel SD implementation:
the master contains the SD search tree and the workers use the 
K-best algorithm to accelerate its exploration. The resulting
hybrid approach enhances the diversification gain, and therefore, lowers the overall complexity
of our parallel SD algorithm.
Our synergistic hybrid SD-K-best approach permits to scale up large MIMO 
configurations up to $100 \times 100$ using modulations with dense constellations, without
sacrificing the BER and complexity. To our knowledge, this is the first
time near-optimal results are reported on such a MIMO dimension in the literature.

\end{abstract}


\section{Introduction}
\label{sec:intro}
Multiple-Input Multiple-Output (MIMO) technology represents a generalization of 
Single-Input Single-Output (SISO) technology that increases the capacity of a radio link by
sending multiple data streams at the same time \cite{telatar1999capacity,foschini1996layered,paulraj1994increasing,raleigh1998spatio}.
Due to their obvious advantages, MIMO systems have already been incorporated into many 
wireless communication network protocols\cite{dahlman20103g,mccann2014official} 
such as IEEE 802.11n (Wi-Fi), IEEE 802.11ac (Wi-Fi), etc.  
Massive MIMO is a new emerging technology that aims to amplify all the benefits of a 
traditional MIMO by further scaling the number of antennas up to several hundreds. 
With the challenge of reaching ten gigabits speed in 5G communication networks and the advent of the 
Internet of Things (IoT)~\footnote{a technology by which various devices, 
domestic and industrial appliances are equipped with an IP address and 
incorporated into a wireless network of "things".}, 
massive MIMO systems are viewed by many researchers and industrials as one of the key technologies to 
sustain a high quality of service when dealing with next generation networks~\cite{mimo-iot}. 
The challenge in these networks resides in the huge number of connected devices, exchanging enormous quantities 
of data (voice, video, etc.) under a real-time response constraint. 
In addition to this challenge, increasing the number of antennas raises several problems, especially in 
terms of energy efficiency and complexity caused by the signal decoding procedure.
Indeed, when scaling up the number of antennas, decoding a message becomes one of the most time-consuming 
operations. In order to maintain a real-time response, researchers generally use linear 
decoders, which are characterized by low complexity with a real-time response, but poor 
performance in terms of Bit Error Rate (BER). 
In order to achieve near-optimal signal decoding, researchers rely on the Maximum Likelihood (ML)
and Sphere Decoder (SD) algorithms \cite{fincke1985improved,viterbo1999universal,hassibi2005sphere}.
The ML decoder performs a brute force exploration of all possible combinations
in the search space of the transmitted vector.
Its complexity increases exponentially with the number of antennas making it impossible, 
in practice, to deploy for massive MIMO systems. 
The SD algorithm is another near-optimal decoder derived from the ML that reduces the size of its search space,
thus, lowering its complexity. Indeed, the SD algorithm compares only the received vector 
with those solutions inside a sphere of a given radius. The radius of the sphere impacts the
complexity and the BER of the overall MIMO system: the smaller the radius,  
the lower the search space (i.e., the complexity), but at the cost 
of possibly missing the actual sent vector if the radius is too small. Tuning the
radius is paramount not only to identify the actual sent vector, but also to 
execute the corresponding procedure under real-time constraints. 
Nevertheless, it turns out that for massive MIMO systems, 
the resulting search space may still be too large to operate on and may engender high complexity.

In this paper, we address the massive MIMO scaling challenge by adapting the SD
algorithm to achieve both decent BER performance and acceptable time complexity. 
To this aim, our contributions are centered around the following three levels. 

The first level of our contributions focuses on revisiting the SD sequential algorithm 
and optimizing the time complexity of its main components. 
The SD algorithm operates on a search tree, where leaf nodes represent 
all possible combinations of the transmitted vector. 
Its goal is to find the combination (leaf node) with the minimum distance from the received signal. 
Two essential aspects of this algorithm must be taken into consideration: 
(1) how to efficiently explore the search tree, i.e., which node to select first, and
(2) how to optimize the evaluation process, i.e., the process of computing the distance 
of each search tree node from the received signal. 
We assess the critical impact of different exploration 
strategies on the complexity of the SD algorithm, namely:  
Breadth-First Strategy (BFS), Depth-First Strategy (DFS), and Best-First Strategy (Best-FS).
We further reduce the complexity of the SD algorithm by reformulating the evaluation process
in terms of matrix algebra to increase the arithmetic intensity. 
We additionally introduce an incremental evaluation
in order to avoid redundant computations. The idea here is to compute the
evaluation of a current node by reusing the evaluations of its previous parent node. 
By choosing Best-FS as the optimal exploration strategy and performing these two aforementioned optimization techniques, 
we significantly reduce the complexity of the sequential SD algorithm, while maintaining an optimal error rate performance. 

The second level of our contributions focuses on accelerating the sequential SD algorithm by using
parallel multicore CPU architectures. 
Our proposed parallel implementation relies on the master/worker paradigm.
It exploits the fact that each path in the SD
search tree can be explored in an embarrassingly parallel fashion.
Indeed, the search tree may be recursively divided into several smaller
search trees where each one is explored by an instance of SD. 
Several instances of the SD algorithm may simultaneously explore the search tree, i.e., 
one instance of the SD algorithm operating as a master process and the others as workers.  
This parallel version aims to diversify the search process, which may rapidly reduce
the radius and thus, the complexity. This method, called \textit{diversification gain},
allows to avoid the exploration of a huge number of branches explored in the serial version. 
However, due to the irregular workload on each path, the parallel implementation 
may run into a load balancing problem, which may affect its parallel scalability.
To overcome this drawback, we propose an efficient dynamic load balancing strategy,
which adjusts the workload per thread at runtime. 
Our proposed parallel approach using our load balancing strategy reports
more than $5\times$ speedup compared to a recent work from Nikitopoulos et al.~\cite{nikitopoulos2018massively} on a 
similar $10 \times 10$ 16-QAM MIMO configuration. It also achieves
up to $60\times$ speedup compared to our serial SD version using a 16-QAM modulation
on a two-socket 10-core Intel Ivy Bridge shared-memory platform (i.e., $20$ cores total).
This represents a super-linear speedup, which has been possible thanks to the diversification gain. 
It turns out that even when using parallelism, the complexity of our SD algorithm may still be
very high to deal with larger MIMO systems and constellation sizes.

To further reduce the complexity, the third level of our work involves 
a trade-off between the complexity and the performance, via a new hybrid implementation 
combining the strengths of our parallel SD and the K-best algorithms.
The main idea of this new implementation, code-named SD-K-best, is to accelerate 
the exploration of the SD search tree stored on the master process 
by using several workers with the low-complexity K-best algorithm. This
approximate method permits to explore rapidly
and partially the subtree sent by the master,  
which reduces effectively the complexity. 
The selected nodes (i.e., branches/paths) are chosen according to their partial distance from the received signal. 
Thus, they are more likely to contain good solutions and may eventually ensure a satisfactory BER. 
Our synergistic SD-K-best implementation integrates all benefits of the parallel SD algorithm 
(i.e., diversification gain, Best-FS, and sphere radius) 
to increase the chances of encountering good combinations of the transmitted signal, 
while reducing effectively the complexity using the parallel SD implementation associated 
with the K-best algorithmic strengths.
The obtained results of our SD-K-best implementation show an 
overall low complexity and good performance in terms of BER, 
as compared to the reference K-best algorithm.
Indeed, for a $16 \times 16$ MIMO system using 64-QAM modulation, 
our SD-K-best approach reaches acceptable error rate at a $20$ dB Signal-to-Noise Ratio (SNR) and real-time 
requirement (i.e., $10$ ms) starting from $28$ dB.  
Last but not least, our SD-K-best approach shows a strong scalability potential by reporting acceptable complexity 
and good error rate performance for a $100 \times 100$ MIMO system using 64-QAM modulation. 
To our knowledge, such a record has never been achieved previously in the literature.

The remainder of the paper is organized as follows. 
Section \ref{sec:rw} summarizes literature on solving massive MIMO systems.
Section \ref{sec:model} describes the system model, recalls the components of the conventional SD algorithm,
and details its exploration strategies. Section \ref{sec:optimization} presents our proposed SD algorithm with 
a new exploration strategy and introduces new optimization techniques. 
Details of our serial and parallel multicore implementations are shown in Section \ref{sec:parallel-SD}. 
Section \ref{sec:hybrid-SD} highlights our new hybrid SD-K-best implementation, which is necessary to tackle massive MIMO systems.
Results and discussions about the trade-off between complexity and performance are given in Section \ref{sec:results}. 
Finally, Section \ref{sec:summary} concludes this paper and summarizes our future perspectives.

\section{Related Work}
\label{sec:rw}
In recent years, there has been a significant body of work 
dealing with massive 
MIMO systems due to their important role in next generation networks. Indeed, 
many surveys have been proposed in the literature highlighting open challenges and recent 
advances in this area \cite{su2013investigation,lu2014overview,marzetta2015massive,bjornson2016massive}. 


Signal decoding in massive MIMO represents the most challenging and critical 
task since the performance of the whole system, in terms of Bit Error Rate (BER), depends on it. 
Signal decoding consists in estimating the transmitted vector by taking into 
account the received vector, which is subject to noise. 
Two kinds of decoders exist in the literature: linear and non-linear (near-optimal) decoders. 
Due to the complexity of non-linear decoders, there are only few works 
in the literature that actually deal with massive MIMO systems. 
Most of these works explore partially the Maximum Likelihood (ML) search tree using 
the Sphere Decoder (SD) algorithm
and/or leverage high performance computing architectures (e.g., GPUs) 
to accelerate the search process and to increase the throughput. 

In \cite{roger2012fully}, Roger et al. propose a parallel fixed complexity SD
for MIMO systems with bit-interleaved coded modulation.  
Their parallel approach exploits multicore processors to
compute the preprocessing phase of the algorithm, 
and the massively GPU hardware resources to process simultaneously 
the detection phase for all $N$ sub-carriers in the system. 

In \cite{jozsa2013new}, Jozsa et al. propose a GPU-based SD algorithm
for multichannel (i.e., sub-carriers) MIMO systems. 
Their approach performs multiple detections simultaneously on the GPU, 
which increases the throughput. 
Moreover, a second level of parallelism introduced within each detection 
relies on the GPU thread block to
accelerate the exploration process of the SD algorithm. 

In \cite{wu2014gpu}, Wu et al. propose an improved version of their initial parallel 
decoder~\cite{wu2012flexible} to increase the throughput of a flexible $N$-way MIMO 
detector using GPU-based computations. This problem consists in dividing the available 
bandwidth into multiple sub-carriers. Each sub-carrier corresponds to an independent
MIMO detection problem. Therefore, the receiver needs to perform multiple MIMO detection procedures.
The authors' idea is to use multiple GPU blocks to execute
multiple MIMO detection algorithms simultaneously. 
To support multiple detections on the GPU, the authors use a
soft-output MIMO detection, which engenders a low memory footprint. 
The results show a good throughput, outperforming the results presented in \cite{roger2012fully}. 

The main problems with the above approaches are twofold. The scalability is a serious bottleneck for large numbers of 
antennas due the limited amount of GPU memory in presence of multi-carriers. Moreover, the high latency
increases the complexity due to the slow PCIe interconnect, when
performing data movement between CPU host and GPU device. 


In \cite{chen2015gpu}, Chen and Leib propose a GPU-based Fixed Complexity 
Sphere Decoder (FCSD) for large-scale MIMO uplink systems. 
The authors reported a speedup around $7\times$ for large MIMO systems and 
constellation sizes compared to their CPU implementation. However, the time complexity 
of their approach is significant even for small numbers of antennas. 

In \cite{arfaoui2016}, Arfaoui et al. propose a GPU-based SD algorithm in 
which a Breadth-First exploration Strategy (BFS) is used to increase the GPU resource occupancy. 
However, increasing the GPU hardware utilization using BFS increases  
the complexity due to the limited impact of pruning process, especially in low Signal-to-Noise Ratio (SNR). 
Our optimized sequential SD implementation herein achieves up to $255$-fold speedup 
on a similar $25 \times 25$ MIMO system with BPSK constellations.

In \cite{husmann2017flexcore}, Christopher et al. 
propose a parallel flexible decoder for large MIMO systems using GPU and FPGA architectures. 
Their algorithm contains two phases. A first preprocessing phase
chooses parts of the SD search tree to explore, and 
a second phase maps each of the chosen parts of the SD tree
to a single processing element (GPU or FPGA).   
The results are presented for a $12 \times 12$ MIMO system using a 64-QAM modulation. 

In \cite{nikitopoulos2018massively}, the authors propose the design and implementation 
of a parallel multi-search SD approach for large MIMO search tree using
multicore CPU and Very-Large-Scale Integration (VLSI) architectures.  
After the preprocessing phase in which they obtain a processing order of the tree branches, 
the authors split the search tree into several sub-trees.
Each sub-tree is then mapped on a processing element and explored using a depth-first strategy.
However, the authors do not take into consideration the load balancing problem, which may arise
in modulations with dense constellations. They also do not update the sphere 
radius at runtime, which may negatively affect the time complexity of their parallel implementation. 
The authors report optimal results for a $10 \times 10$ MIMO system using  
16-QAM modulation and approximate results for a $16 \times 16$ MIMO system using 64-QAM modulation.

Most of the existing works report experimental results
for rather small MIMO configuration systems and do not report or 
satisfy the real-time response constraint.  
In addition, they rely on GPUs to accelerate
the partial or complete exploration of SD search-trees. While GPUs
are throughput-oriented devices, the resulting size of the SD search space
still remains prohibitive to maintain a decent time complexity. 

We decide herein to revisit the fundamentals of the popular serial SD algorithm, 
which stands as a proxy for all non-linear decoders. We reduce its time complexity by
relying on a new Best-First Strategy (Best-FS) for efficient exploration, a matrix algebra reformulation for
increasing arithmetic intensity, and an incremental evaluation process for cutting down the number of operations.
These optimizations are performed for all SNR regions allowing to reduce the time compexity,  while maintaining optimal BER performance.
We then extend the sequential implementation by exploiting the inherent
parallelism of the SD algorithm. We take advantage of the diversification gain 
to avoid the exploration of a huge number of branches explored in the serial version. 
We employ a dynamic load balancing 
scheduler that minimizes idleness, communications, and synchronization overhead.
Finally, in order to break the symbolic barrier of hundreds of antennas for the first time, 
we deploy a new hybrid approximate approach that blends the aforementioned strengths of our SD
implementation with the ones from the K-best algorithm. 
This new SD-K-best CPU-based implementation achieves performance and complexity metrics
at unprecedented levels from the literature, even with GPU hardware accelerators.
\section{System Model and Conventional SD Algorithm}
\label{sec:model}
\subsection{System Model}
\begin{figure}[t]
 \centering
\includegraphics[width =9.75cm]{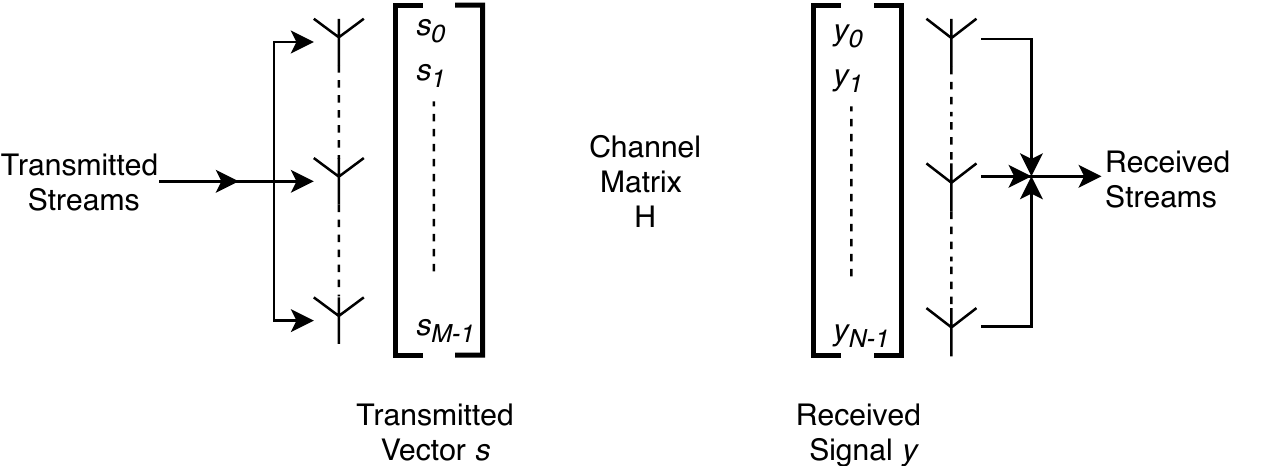}
\caption{
Example of a MIMO system where the vector $\bm{s}$ is transmitted by  $M$ 
transmitter antennas via a channel matrix $\bm{H}$. 
The received vector $\bm{y}$ is a collection of  $N$ receiver antennas' observations.  }
\label{fig:11}       
\end{figure}

In this paper, we consider a baseband MIMO system consisting of $M$ transmit 
antennas and $N$ receive antennas, as depicted in Figure \ref{fig:11}. 
The transmitter sends $M$ data streams simultaneously to a receiver
using multiple antennas via a flat-fading channel. 
This system is described by the input-output relation in the following Equation \ref{eq:1}:
\begin{eqnarray} 
\label{eq:1}
\bm{y}=\bm{Hs + n},\\
\nonumber 
\end{eqnarray}
where the vector $\bm{y}=[y_1,...,y_N]^T$ represents the received signal.
$\bm{H}$ is an $N \times M$ channel matrix, where each element $\bm{h_{ij}}$ 
is a complex Gaussian random variable that models the fading gain between 
the $j$-th transmitter and $i$-th receiver. The vector $\bm{s}=[s_1,...,s_M]$  
represents the transmitted vector, where $\bm{s_i}$ belongs to a finite 
alphabet set denoted by $\Omega$. 
Finally, $\bm{n}=[n_1,...,n_N]^T$ represents the additive white Gaussian 
noise with zero mean and covariance $\bm{I}_{N}$, where $\bm{I}_{N}$ designates 
the identity matrix of size $N$.  
For convenience, let us consider $\bm{S}$ as the set of all possible 
combinations  of the transmitted vector $\bm{s}$. 
The possible number of combinations corresponds to the complexity of the MIMO system 
and it is calculated as follows: $|\bm{S}|=|\Omega|^M$. 


There are two options to decode the received signal. Either we use linear 
decoders characterized by low complexity and poor performance in terms of 
BER, or we use non-linear (optimal) decoders characterized by good BER quality but high complexity. 

Linear decoders multiply and map the received signal using a matrix denoted 
by $\bm{H_{inv}}$ ($M\times N$), obtained from  the channel matrix  $\bm{H}$.
The most commonly used linear decoders in the literature define $\bm{H_{inv}}$ as follows: 
\begin{itemize}
\item Maximum Ratio Combining (MRC), where 
the $H_{inv}$ is equal to:  $$\bm{H_{inv} = H^{H}}.$$
\item Zero Forcing (ZF), where 
the  $H_{inv}$ in case of  $M \leq N$ is equal to:
$$\bm{H_{inv}= (H^H \cdot  H)^{-1}\cdot H ^{H}}.$$ 
\item  Minimum Mean Square Error (MMSE), where 
the $H_{inv}$ matrix used by this decoder is equal to:
$$\bm{H_{inv}}=   (\bm{H^H \cdot H }+ \frac{1}{SNR}\bm{ \cdot \bm{ I_m})\cdot H ^{H}},$$
with the Signal-to-Noise Ratio $SNR=P$, 
where \textit{P} is the average transmit power, since we normalize the
noise covariance to identity, without loss of generality.
\end{itemize}

As for non-linear decoders, the Maximum Likelihood (ML) is the \emph{de facto} decoder, exhibiting 
high complexity. 
It calculates  \textit{a posteriori} probability for each possible 
transmitted vector $\bm{s\in S}$.  
In other words, the algorithm performs a brute-force exploration of the 
entire search space, as shown in the following Equation \ref{eq:2}:
\begin{eqnarray} 
\label{eq:2}
\bm{\hat{s}_{ML} = arg\min_{s\in S}  ||y - Hs||^2.}\\
\nonumber 
\end{eqnarray}

The ML decoder chooses the vector \textit{s} 
that minimizes the distance between the received vector \textit{y} and the 
assumed transmitted vector $\bm{Hs}$. 
In perfect conditions, i.e., in absence of noise, this minimum distance is equal to zero, which indicates that 
the transmitted vector is exactly the received one, up to a channel multiplication. 
For more details about the ML decoder, the reader may refer to \cite{Marvin-book}.
Another example of non-linear decoders is the Sphere Decoder (SD) algorithm 
\cite{viterbo1999universal,agrell2002closest}. 
This latter mimics the ML decoder, but limits the search for the candidate 
vector to a smaller space than ML, reducing enormously the complexity. 
The SD algorithm consists in exploring solutions inside a 
sphere of radius $r$ set initially by the user, as shown in the following Equation \ref{eq:3}:
\begin{eqnarray}
\label{eq:3}
  \bm{  ||y - Hs||^2} <r^2, \ \ where \ \bm{ s\in S.}\\
\nonumber
\end{eqnarray}
The radius may then be updated subsequently during the search process
at runtime to further prune the search space and reduce the complexity.
In the following section, we describe the baseline SD algorithm and its components in more details.


\subsection{Conventional Sphere Decoder Algorithm} 
The SD algorithm operates on a search tree that models all possible  
combinations of the transmitted vector. 
This algorithm aims to find the best path in terms of distance from the 
received signal, while ignoring non promising branches.
Equation \ref{eq:3} can be translated in solving the integer least-square problem. 
It starts with a preprocessing operation by performing
a $\bm{QR}$ decomposition of the channel matrix $\bm{H}$ as $\bm{H=QR}$, 
where $\bm{Q \in \mathbb{C}^{N\times N}}$ is an orthogonal matrix 
and $\bm{R \in \mathbb{C}^{N \times M}}$ is an upper triangular 
matrix. 
This preprocessing step permits to expose the matrix structures of $\bm{Q}$
and $\bm{R}$, which will be eventually used to simplify the computations.
Indeed, by using the orthogonality of $\bm{Q}$ and considering 
only the ${M\times M}$ upper part of $\bm{R}$, the problem defined in the Equation \ref{eq:3} 
can be transformed into another equivalent problem as follows: 

$$\bm{||y - Hs||^2 = ||y - QRs||^2}$$
$$=\bm{||Q(Q^Hy - Rs)||^2$$
$$ = ||Q^Hy - Rs||^2} $$
$$= ||\bm{\bar{y} - Rs||^2},\ where\ \bm{\bar{y} = Q^Hy}$$
$$= ||\ \ 
 \begin{bmatrix}
\bm{     \bar{y}_0 }\\
     \bm{\bar{y}_1}\\
     .\\
    .\\
    .\\
   \bm{  \bar{y}_{M-1}}\\
      \end{bmatrix}
      -
  \begin{bmatrix}
   r_{00} &r_{01} & ....& & r_{0M-1} \\
    0 &r_{11} & ....& & r_{1M-1}\\
    . &.& ....& & .\\
    . & .& ....& &.\\
     . &  .& ....& &.\\
  0 &0&...& 0 & r_{M-1 M-1}\\
      \end{bmatrix}
       \begin{bmatrix}
   \bm{   s_{0} }\\
    \bm{ s_{1}}\\
     .\\
    .\\
    .\\
 \bm{    s_{M-1}}\\
      \end{bmatrix}
   \ \ ||^2.
$$
Therefore, finding the supposed transmitted vector ($\hat{s}$) in Equation \ref{eq:1} 
is equivalent to solving the following minimization problem: 
\begin{eqnarray}
\label{eq:4}
\min \sum_{k=1}^{M} g_k(\bm{s_{M-1},...,s_{M-k}}), \\
\nonumber 
\end{eqnarray} 
where $g_k(\bm{s_{M-1},...,s_{M-k}})=  ||\bm{\bar{y}_{M-k}}-\sum_{i=M-k}^{M-1} (r_{(M-k),i} \bm{s_{i}}) ||^2.$
This latter formulation of the problem allows us to model all  
possible combinations of the transmitted vector (i.e., search space) as a search tree with $M$ layers.  
To find the path with the minimum distance from the received signal, the SD 
algorithm is decomposed into three components: branching, evaluation, and pruning. 
\begin{figure*}

\centering

\includegraphics[width =15cm]{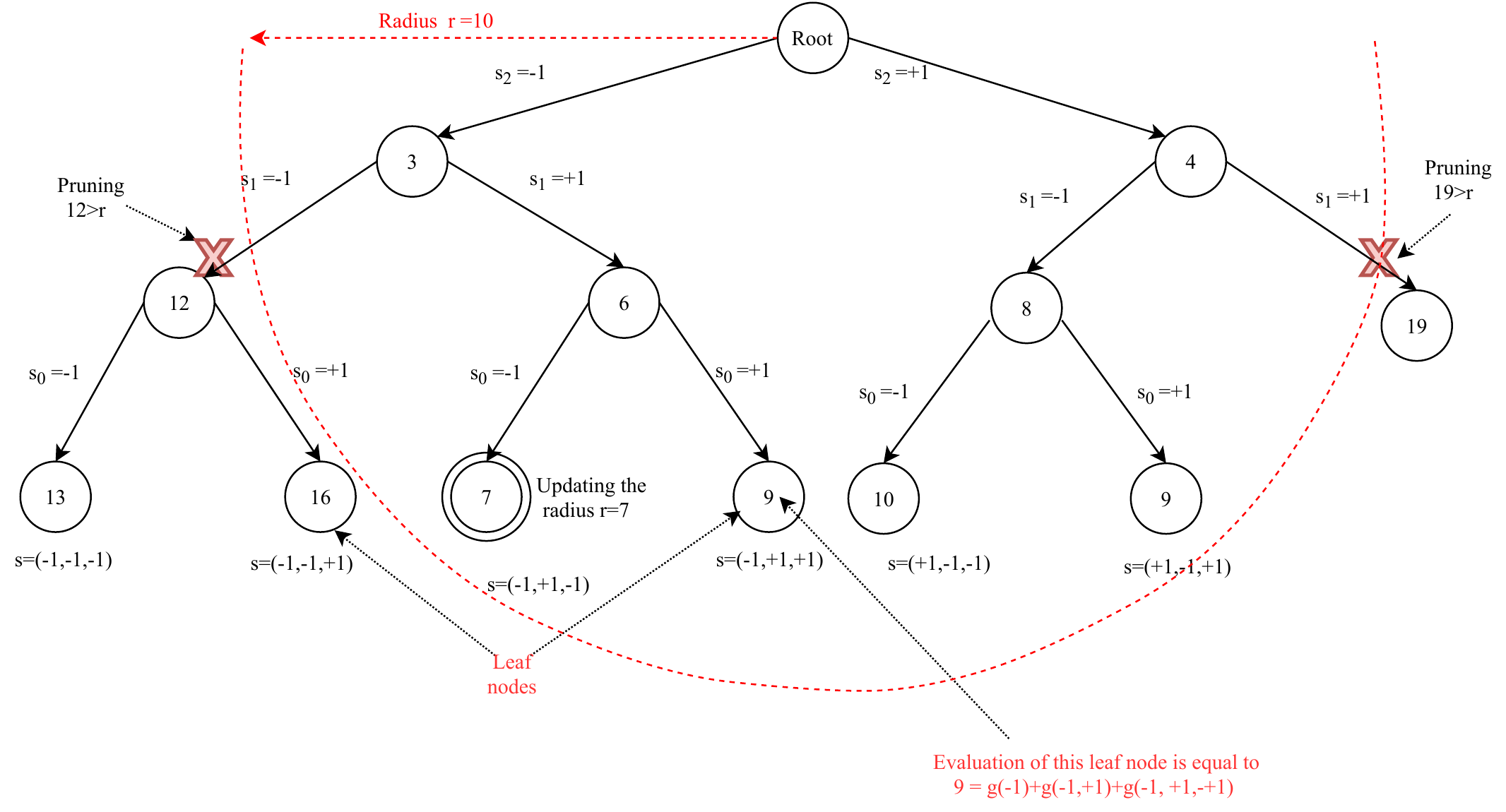}
\caption{
This figure represents an example of the SD search tree for a MIMO system with 
three transmit antennas.  
At each level, one symbol is fixed, starting with the last one. 
The partial evaluation of each node (sub-problem) is stored inside the 
circles.  
The pruning process uses the evaluation of the node and the sphere radius 
\textit{r} to avoid unpromising branches.}
  \label{fig:1}    
\end{figure*}
\begin{algorithm}
\caption{The Sphere Decoder (SD) Algorithm.}
\label{alg:1}
\SetAlgoLined
\KwData{Received signal $\bm{y}$\\
Constellation order $\Omega $\\
Channel estimation $\bm{H}$\\
Noise variance estimation $\sigma^2 $\\
Radius $\bm{r}$
}
\KwResult{Decoded vector $\bm{\hat{s}}$}
 initialization\;
 
 List $<--$  root\;

 \While{List != $\emptyset $}{
  \textit{P} = select\_node (List)\; \label{line:exploration}
  
  List = List - \{\textit{P}\}\;
  
  Generate successors $P_{i}\  of\  P\  /\ i=\{1,...,|\Omega |\}$\; 
  
  \For{\textbf{each} $P_{i}$}
  {
    \eIf{$E(P_{i})<r$}{
    \eIf{$P_{i}$ is a leaf node (complete solution)}{ 
   $r$= $E(P_{i})$\; (The evaluation of sub-problem $P_{i}$.)
   
   $\hat{s}$ = $F_{P_{i}}$ \;
   }{
   List = List $\cup\ P_{i} $\;
  }
  }{prune the branch\;}
 }}
\end{algorithm}
Figure \ref{fig:1} shows an example of the SD search tree
and highlights the SD components on a MIMO 
system with three transit antennas using the Binary Phase-Shift Keying 
(BPSK) modulation. 

In the following, we describe formally each of the SD components, as described 
in Algorithm \ref{alg:1}.
\subsubsection{Branching}
the branching component for a MIMO system with \textit{M} transmit antennas is performed 
over the symbols of a transmitted vector. This process creates a search tree 
with \textit{M} levels, so that each level corresponds to one symbol. 
Thereby, the last level of the search tree contains all possible combinations
of the transmitted vector. 
Each search tree node is characterized by a set of fixed symbols denoted by $F$.
In this way, there is no fixed symbols in the root node  
($F_{root}=\emptyset$). 
The branching component is essentially a recursive process that divides the search space related to a 
search tree node $P$  over several successors (or sub-problems) $P_i, i=1,...,|\Omega|$. 
Each subsequent successor is eventually handled in the same way until a 
complete solution is found, i.e., until the number of symbols in the 
solution is equal to $M$. 
The number of immediate successors depends on the size of the alphabet. 
For example in Figure \ref{fig:1}, the size of the constellation in the 
BPSK modulation is two (-1 and +1). In this case, we have two immediate 
successors ($P_1$,$P_2$) of node \textit{P}, where $P_1$ is characterized by 
the set $F_{P1}$=$F_p \cup \{-1\}$ and $P_2$ is characterized by the set $F_{P2
}$=$F_p \cup \{+1\}$.  
At each level of the search tree, we perform branching over one symbol.
Since matrix $\bm{R}$ from the $\bm{QR}$  decomposition is upper triangular, 
each level \textit{l} of the search tree corresponds to a symbol $\bm{s_{M-l}}$. 
For example, level 1 corresponds to symbol $\bm{s_{M-1}}$ and level \textit{M} 
corresponds to symbol $\bm{s_{0}}$. Thus, we begin by fixing $\bm{s_{M-1}}$, 
then symbol $\bm{s_{M-2}}$, and so on, until we reach the leaf nodes where
symbol $\bm{s_{0}}$ is fixed.
   
\subsubsection{Evaluation}
the evaluation component represents the process of computing the 
Partial Distance (PD) of each search tree node from the received signal.
After the branching process, the evaluation, denoted by \textit{E} in Algorithm \ref{alg:1}, 
is calculated for each successor using Equation \ref{eq:4}.  
More precisely, the evaluation of a search tree node \textit{P} characterized by 
\textit{L} fixed symbols ( $F_p=\bm{\{s_{M-1}, ...,s_{M-L}\}}$) is defined as
$E(P)= \sum_{k=1}^{L} g_k(\bm{s_{M-1},....,s_{M-k}})$.
This comes down to calculate:  
$$E(P)= || 
 \begin{bmatrix}
   \bm{  \bar{y}_{M-L}} \\
    .\\
    .\\
    \bm{ \bar{y}_{M-1}}\\
      \end{bmatrix}
      -
  \begin{bmatrix}
    0 & .& ....&r_{M-L, M-2} & r_{M-L, M-1}\\
    . & .& ....&. &.\\
    . &  .& ....&. &.\\
    0 &0&...& 0 & r_{M-1,M-1}\\
      \end{bmatrix}
       \begin{bmatrix}
   \bm{   s_{M-L}} \\
    .\\
    .\\
    \bm{ s_{M-1}}\\
      \end{bmatrix}
     ||^2.
$$
This means that we use only the last \textit{L} elements in vector $\bm{\bar{y}}$ 
and the last \textit{L} lines in matrix $\bm{R}$ to compute the evaluation of a 
node \textit{P} with \textit{L} fixed symbols where $L=|F_p|$.  

\subsubsection{Sphere Radius and Pruning}
the sphere radius defines the region of the search space in which an intelligent 
enumeration can be performed. The radius represents an important parameter for 
determining the complexity of the SD algorithm, since a large value of the radius induces a high complexity.  
The ideal value for the radius should as small as possible, but as long as the
corresponding region still includes the ML solution. 
The sphere radius, denoted by $r$, imposes an upper limit for the 
expression $||\bm{\bar{y} - Rs ||^2} $, which leads to reject any partial 
combination of the transmitted vector $\bm{s}$  with a partial evaluation greater than $r$.
The sphere radius used in this paper is equal to $r^2=  N \cdot M \cdot 10 ^{\frac{-SNR}{10}} $
and is the same used in \cite{arfaoui2016}. Tuning and studying the impact 
of $r$ on complexity and performance is critical but beyond the scope 
of this paper. 

The pruning process consists in detecting and eliminating the unpromising 
branches in the search tree by using both the sphere radius and the evaluation of nodes.  
As seen in Equation \ref{eq:4}, the evaluation increases each time we fix a new 
symbol in the transmitted vector. This means that a node $P$ with a partial 
evaluation $E(p)>=r^2$ can not lead to a complete solution that improves the best one already found. 
In this specific case, the node is eliminated. In order to ensure an efficient pruning process 
during the search, we replace the value of the radius each time a new better 
solution (leaf node) $\bm{s\in S}$ is explored, i.e., $r^2 = E(\bm{s})$.  
Updating the value of the radius during the pruning phase is very important for the  
subsequent explorations of the search tree.  This feature may prevent exploring
a huge number of branches that are outside of the sphere radius. 

\begin{figure*}[ht]
\centering
\includegraphics[width =15.2cm]{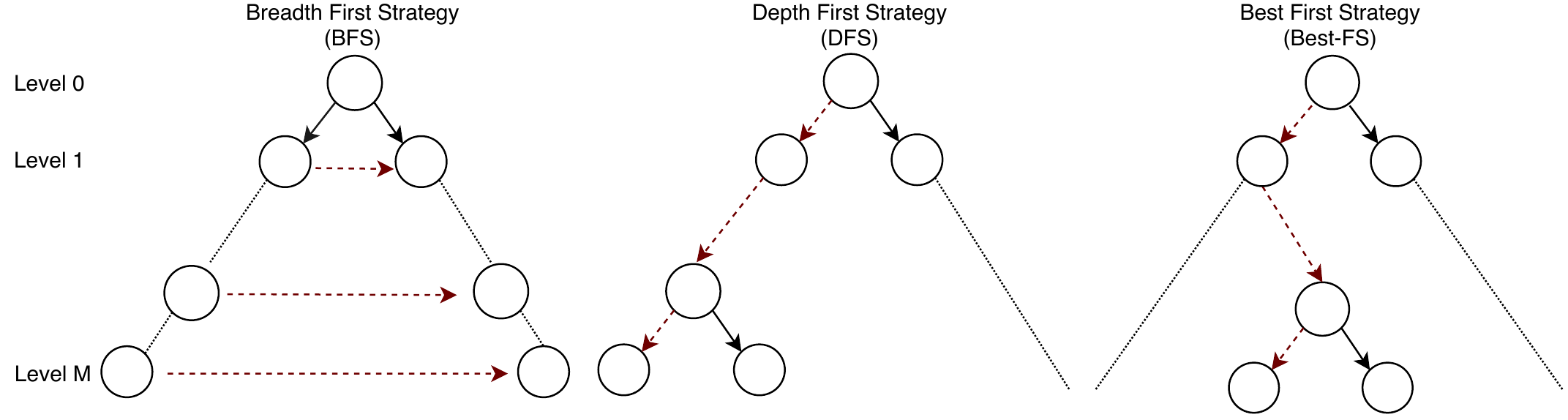}
\centering\caption{Exploration strategies used by the SD algorithm.}
\label{fig:3}       
\end{figure*}

\subsection{Exploration Strategies}
To give an idea about the magnitude of the search space, the number of combinations (leaf nodes) for
a MIMO system with the BPSK modulation and fifty transmit antennas is 1.1258999 $e^{+15}$.
Exploring all these possibilities under real-time constraints is prohibitively expensive.
The exploration strategies for the SD algorithm define the way the search tree is explored and traversed, 
as illustrated in line \ref{line:exploration} of Algorithm \ref{alg:1}.
The performance of each exploration strategy depends on the MIMO configurations and the 
underlying hardware architectures. 
For this reason, we investigate in this paper the impact of several
exploration strategies on the SD complexity. 
The SD search tree is stored using a list structure and is partially explored at each iteration. 
Figure~\ref{fig:3} shows the two typical ways of 
exploring a tree: Breadth-First and Depth-First. 
 
\subsubsection{Breadth-First Traversal Strategy}
the Breadth-First Strategy (BFS) explores the search tree level by level, which means that all nodes of a given 
level must be explored before moving toward the lower levels.
In practice, implementing the BFS consists to apply First-In First-Out (FIFO) 
strategy on the data list that contains the tree, 
i.e., selecting always the rightmost node in the list.  
The BFS is particularly very suitable for parallel implementation since all nodes 
of a given level can be treated independently. This enables to efficiently exploit  
the available computing resources. However, its major drawback is the high memory footprint 
during the search process. 
This makes its application very limited in practice, especially for massive MIMO 
systems where the number of possible solutions may be tremendous. 
The second major drawback of the BFS strategy is the fact that the sphere radius 
remains the same throughout the search process, since this 
strategy reaches the leaf nodes only at the last level. The static sphere radius cannot be updated
at runtime and engenders a poor pruning process, which induces a very high complexity even 
for small MIMO systems.   

\subsubsection{Depth-First Traversal Strategy} 
the Depth-First Strategy (DFS) is a recursive process based on a backtracking technique. Unlike the BFS, 
the DFS aims to reach leaf nodes as quickly as possible by exploring down the 
current path. Once it reaches the leaves, DFS may explore backward to retrieve new nodes
and carry on again along the new path until attaining the bottom of the tree. This
process is pursued until all nodes are explored.
In practice, implementing the DFS consists to apply a Last-In First-Out (LIFO)
strategy to the data list that contains the tree. 
In other words, the DFS always select the leftmost node in the 
list, which is the most recently added to the list after the branching process. 
The interesting fact that should be highlighted for DFS in 
general is the limited memory usage. This feature makes it very 
suitable in practice for challenging problems such as decoding messages in massive MIMO systems.  
The other interesting fact is the possibility of updating the value of the sphere 
radius dynamically due to the huge number of entirely explored solutions. 
Although the complexity may be improved for sequential implementations, the DFS does not expose 
parallelism compared to BFS. Therefore, it may not be suitable in presence of high
number of computing resources needed to operate large MIMO
configurations. 

\section{Leveraging The Sphere Decoder Algorithm Toward Massive MIMO}
\label{sec:optimization}
In this section, we present an initial set of optimization techniques to improve
the exploration and the evaluation phases in our SD algorithm.

\subsection{Best-First Strategy}
In order to improve the search tree exploration,
we introduce the Best-First Strategy (Best-FS).
This strategy is very similar to the DFS since both are meant to explore leaf nodes first. 
However, the Best-FS targets a better quality of leaf nodes (in terms of
distance from the received signal) as compared to the DFS exploration model. 
After the branching process, Best-FS chooses first the node with the best evaluation 
in order to complete its exploration.
The only difference against the DFS model is that the nodes generated after the 
branching process are sorted according to their partial distance before being inserted into the list. 
Since the number of nodes generated after the branching is limited, the overhead time of
the sorting process is insignificant. 
The exploration based on Best-FS is theoretically
more suited for SD implementation since it targets better quality leaf-nodes. 
Therefore, this approach proactively reduces the sphere 
radius throughout during the SD process, which decreases the number of explored nodes and thus, the 
memory footprint and the arithmetic complexity.

After optimizing the exploration phase, we aim to further 
reduce the SD complexity by optimizing its evaluation phase. 
This latter represents the most time-consuming part of the SD algorithm,
since it is calculated for each search tree node. 
To achieve this goal, we consider two aspects: reducing the number of 
evaluation steps and avoiding redundancy in the evaluation process.

\subsection{Grouping Evaluation Steps}
The idea here is to reduce the number of intermediate evaluation points for each path in the search tree.  
For a MIMO system with $M$ transmit antennas, we generally perform $M$ 
evaluation points to reach the leaf nodes, which may be overwhelming when 
scaling up massive MIMO systems.    
Reducing the number of evaluation points can be achieved by performing the branching 
process simultaneously over several symbols instead of one at a time. 
Indeed, performing 
the branching over $J$ positions in the transmitted vector allows us to reduce the 
number of evaluation points from $M$ to $M/J$. 
For instance, performing the branching over five symbols for a MIMO system with $100$
transmit antennas will reduce the number of evaluation points for each search tree
path from $100$ to only $20$. 
Beside shortening the overall processing time of the evaluation phase,
this grouping technique allows to reach the leaf nodes more quickly which may result 
not only in reducing the latency overhead but also in pruning 
earlier a lot of branches.  

However, we should keep in mind that the number of immediate successors will increase 
according to the number of fixed symbols in the branching process. 
Therefore, instead of creating $|\Omega|$ new successors, we create $|\Omega|^J$ new 
successors. The parameter $J$ should be tuned accordingly to trade-off complexity and  
parallelism.


\subsection{Incremental Evaluation}
Since the search tree for massive MIMO may be huge, it is paramount to optimize
the evaluation in order to achieve good performance.
Our goal here is to further reduce the complexity of the evaluation step by avoiding 
redundant computations.  

We recall that the evaluation of a search tree node $P$ with $L$ 
fixed symbols is equal to $E(P)= \sum_{k=1}^{L} g_k(\bm{s_{M-1},..., s_{M-k})}$.
We can see that the complexity of the evaluation increases 
significantly when moving toward leaf nodes.
In order to avoid this increase in complexity and to have the same evaluation time for 
all search tree nodes, we take advantage of the incremental nature of the evaluation 
process for this problem. 
Indeed, the evaluation of successors $P_i$ of the node $P$ with $L_i$ fixed symbols, 
where $L_i>L$, can be decomposed as follows:
\begin{eqnarray}
\label{eq:inc}
E(P_i)= \sum_{k=1}^{L_i} g_k(\bm{s_{M-1},..., s_{M-k})} =
  \overbrace{
   \underbrace{\sum_{k=1}^{L} g_k(\bm{s_{M-1},..., s_{M-k})}}_\text{E(P)} + \underbrace{\sum_{k=L+1}^{L_i} g_k(\bm{s_{M-1},..., s_{M-k}).}}_\text{non-computed part}
  }
\end{eqnarray}


Therefore, we accumulate the calculations during the evaluation of previous nodes
in order to use it later when evaluating the successors. 
In this way, the evaluation process for all search nodes consists to compute only 
the non-computed part of Equation \ref{eq:inc}.

\section {Implementation Details of The Parallel Sphere Decoder Algorithm} 
\label{sec:parallel-SD}
This section provides details of the sequential and parallel 
Sphere Decoder (SD) implementations based on the Best-FS combined with the grouping
and incremental evaluation steps.

\subsection{Serial Implementation of the SD Algorithm}
In the following, we describe  our idea of optimizing the implementation of the 
evaluation phase using the well-known Basic Linear Algebra Subprograms (BLAS). 

As mentioned earlier, performing the branching over $J$ symbols at a time allows us to 
reduce the number of evaluation steps. However, it also increases the number of 
successors from  $|\Omega|$ to $|\Omega|^J$. 
Each successor is characterized by a vector of fixed symbols called here $v$.
To evaluate all successors at once, we propose the following: We begin by regrouping 
all successors' vectors in one matrix named $V$, with $|v|$ lines (number of fixed 
symbols) and $|\Omega|^J$ columns (number of successors).  
After that, we create a matrix $\bm{R^\prime}$  by considering only the last $|v|$ 
lines  and the last $|v|$  columns of the matrix $\bm{R}$. 
Thereafter, we use the well-known BLAS library \cite{blas} to compute $B= Y^\star- RV$, 
where $Y^\star$ represents the last $|v|$ elements of $\bm{\bar{y}}$ duplicated $|\Omega|^J$ times.  
Finally, we use the matrix $B$ to deduce the evaluation of each successor by computing 
the norm over the columns of matrix $B$.  
In this way, we have been able to optimize the implementation of this algorithm. 

However, searching for the optimal combination of the transmitted vector is a time 
consuming operation due to the large scale of the SD search tree.  
In this section, we study the impact and possible gain of exploiting
the processing elements of a single workstation. 
Indeed, most of today's machines are parallel from a hardware perspective;
offering a decent computing power, which is not exploited in most cases. 
For this reason, and to evaluate the possible gain that can be achieved
using small number of computing resources, two parallel SD approaches are proposed.

\begin{figure}[h]
	\centering
	\includegraphics[width=9cm]{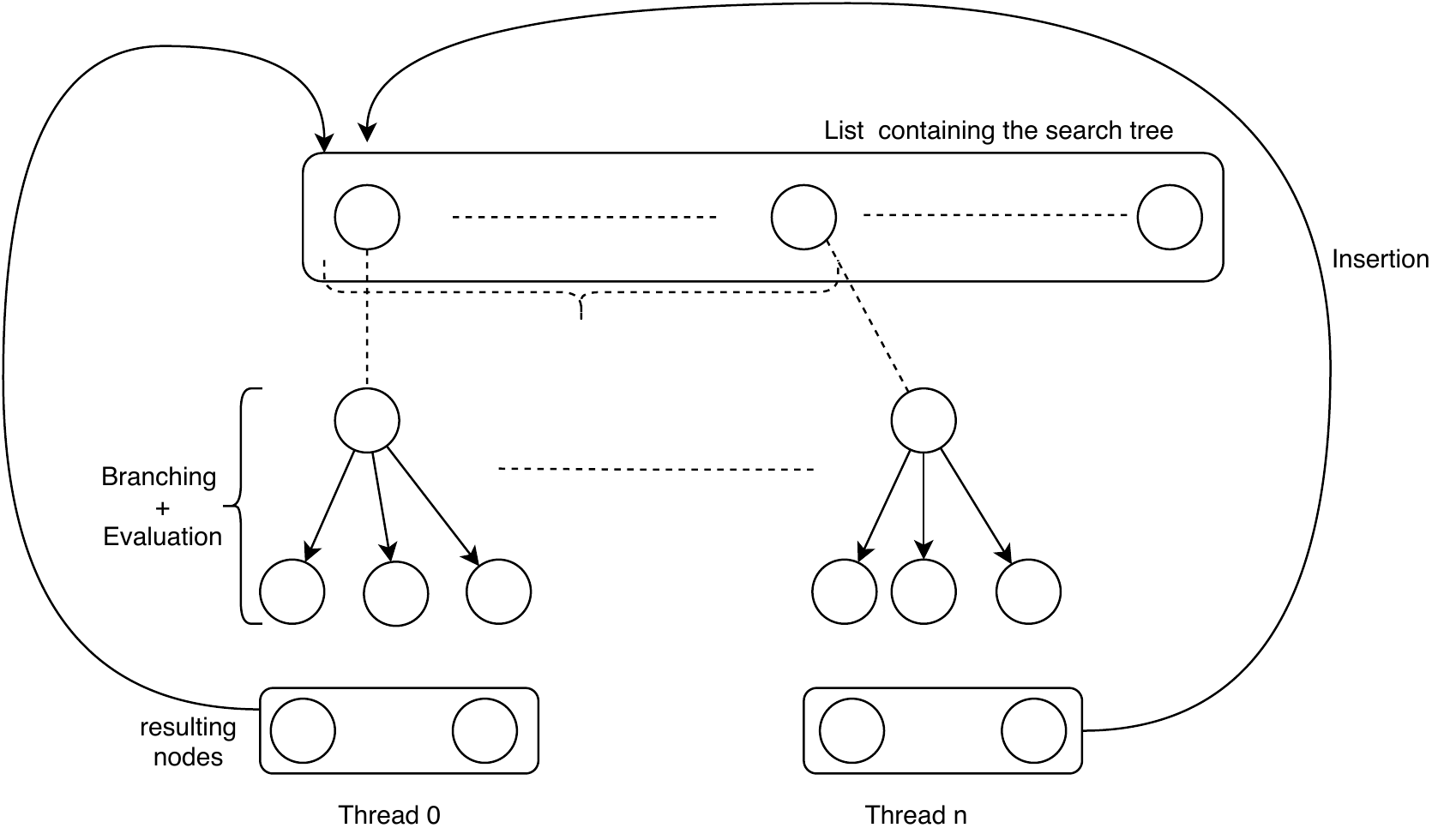}
		\caption{Low-level SD parallelization.}
	\label{fig:p2}
\end{figure}

\subsection{Low-level Parallel SD Approach}
Our first attempt to accelerate the SD algorithm is by accelerating its exploration process. 
As depicted in Figure \ref{fig:p2}, this first approach aims to accelerate
the sequential process of exploring one search tree stored in a list.  
At each iteration,  the serial SD algorithm takes  a search tree node from the list and
performs the branching operation, which creates a set of successor nodes.  
After-that, the SD  calculates the partial distance (evaluation) for all resulting nodes before adding them to the list. 
The idea of this parallel approach is to perform the branching overs several nodes at a time. 
Therefore, at each iteration of the SD algorithm, we create a set of threads to perform the same sequential process 
(branching and evaluation) over several search tree nodes in a concurrent safe way. 
This process is repeated until the list becomes empty. Hence, the end of the parallel algorithm is reached. 
The number of threads and work-load for each one  must be adapted 
to the number of processing elements available in the machine. 
Moreover, to ensure low memory utilization, this approach uses the Best-FS exploration model.

The interesting fact about this approach is the fair work-load distribution between parallel threads, 
which prevents the idleness especially  for this kind of problems.
The down side of this approach is the scalability issue that may occur when increasing the
number of parallel threads, due to the concurrent access to the same data-structure. 
To avoid this problem, a second parallel SD approach is proposed.

\begin{figure}[ht]
 \centering
 \includegraphics[width =9cm]{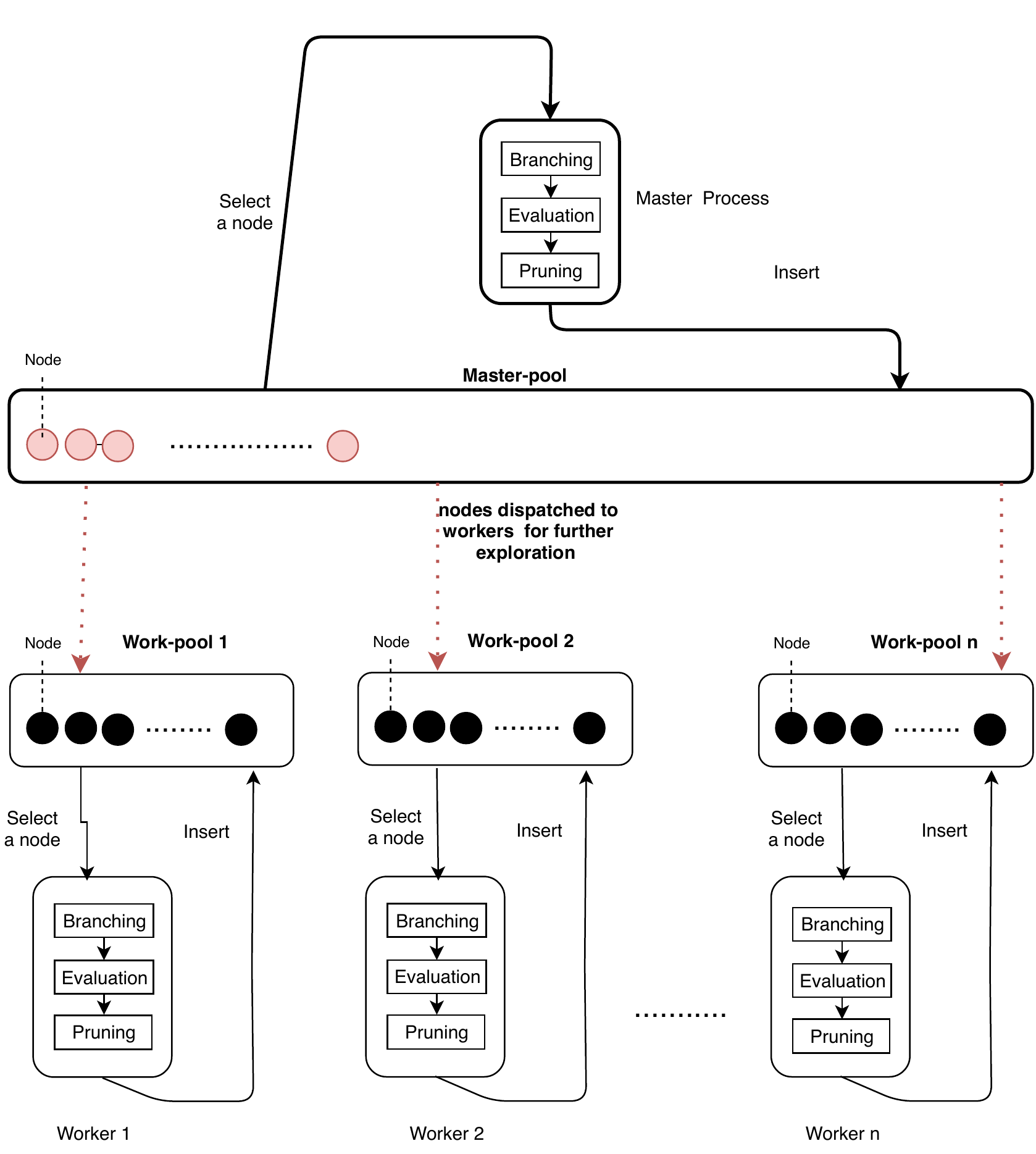}
 \caption{High-level SD parallelization.}
 \label{fig:PSD}       
 \end{figure}
\subsection {High-level Parallel SD Approach}
As depicted in Figure \ref{fig:PSD}, our second parallel scheme can be seen as a high 
level parallelization in which  several instances of the SD algorithm explore  
simultaneously the search space.  
Indeed, this scheme exploits the fact that the global tree which models all possible  
combination of the transmitted vector can be divided into several smaller subtrees 
where each can be explored independently from the others. 
The only shared information between the SD instances is the value of the sphere radius; 
which is updated each time a new better solution is explored by parallel threads.  
The proposed parallel scheme, which exploits the multi-core CPU processors, is based 
on the \textit{Master/Worker} paradigm.
According to this paradigm, we have one instance of the SD algorithm playing the 
role of the master process and other SD instances as workers. 
The master divides the search tree into several sub trees, which are meant 
to be explored by workers. We define a work-pool as a set of active nodes  
generated by the SD algorithm during the search process. Two kind of work-pool can be  
identified: a master-pool owned by the master process, and several local work-pools  
owned by the different workers. 
Initially, all workers are blocked, waiting for nodes to explore. 
The master creates the root node and begins the exploration of the search tree which  
generates a set of nodes in the master-pool. When the number of nodes in the master-pool
 is greater than the number of workers, the master wakes up blocked workers by 
sending to each one of them a node (subtree). After that,  each 
worker launches its own SD instance to explore the sub-tree related to the received node. 
In order to efficiently reduce the sphere radius, all parallel SD instances (threads) explore their 
subsequent subtree according to the Best-FS model.  
The master periodically checks on the state of workers and wakes-up any blocked one 
(worker with an empty work-pool). 
Each time the master-pool is empty, the master checks the state of all  workers. 
If all of them are blocked, the master sends an end signal to all parallel threads. 

\textit{Load-balancing problem}:
Due to the prohibitive complexity of SD, and the irregular work-load in 
the SD sub-trees, a load-balancing strategy must be used. 
This latter has an objective to increase the efficiency of our high-level
parallel SD approach by avoiding the idleness of workers. 
In our case, the idleness of workers appears only in the case where the master work-pool is empty.
Our idea to avoid the  idleness of workers is to perform a workload
redistribution over all blocked workers whenever the master work-pool is empty. 
In this case, the master locates the worker which has the highest number of unexplored nodes. 
Then, it distributes them over blocked workers and move the most of remaining nodes to its own work-pool.
In this way, we have been able to ensure a fair work-load distribution during the decoding process.

The parallelization allows to speedup enormously the exploration process and reduce the SD complexity. 
However, the complexity of the latter is still very high to deal
with massive MIMO systems under real-time constraint. 
For this reason, we propose in the following a new approximate algorithm that perform a 
trade-off between the complexity and the performance in terms of error rate. 

Similarly to the DFS, the Best-FS may be less suitable for parallelization as compared to 
the BFS exploration model. The number of nodes handled at each iteration is limited, leading
to a low hardware occupancy. 

\section{Our Hybrid Sphere Decoder / K-Best Algorithm}
\label{sec:hybrid-SD}
The main idea of our proposed approximate approach is to achieve an acceptable BER in real-time complexity, i.e., 
losing a bit in performance (as compared to the SD algorithm),
 but wining in terms of complexity.  
The challenge is to find the appropriate balance to achieve
 both near ML performance and real-time response. 

\begin{figure}[ht]
 \centering
 \includegraphics[width =8.5cm]{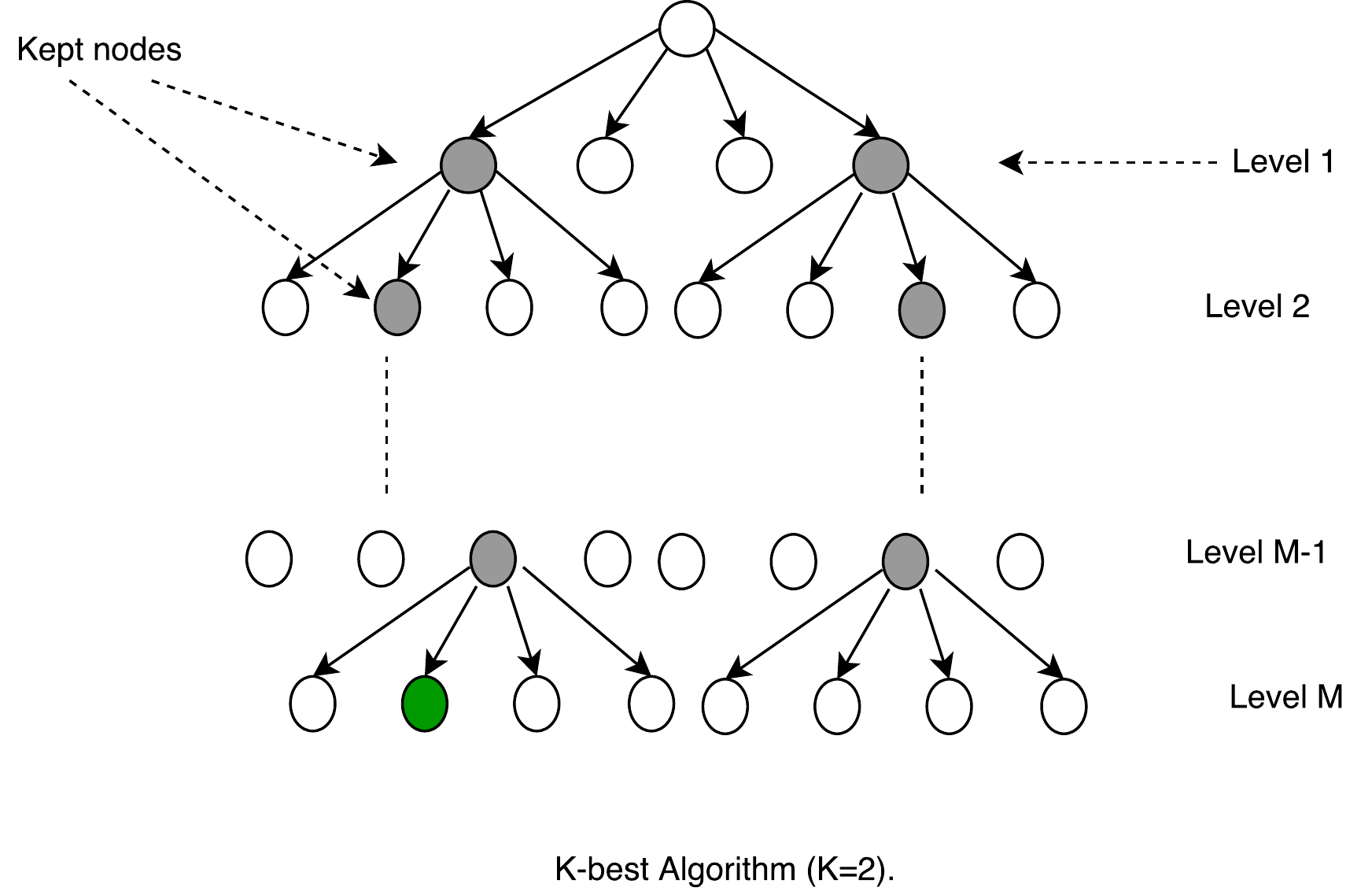}
 \caption{The K-best algorithm with K=2.}
 \label{fig:kbest}       
 \end{figure}

One of the best reference algorithms performing a trade-off between the complexity 
and the performance is the K-best algorithm \cite{wong2002vlsi}. 
Similarly to SD, the K-best algorithm \cite{wong2002vlsi} operates on a search 
tree that models all possible combinations of the transmitted vector (see Figure \ref{fig:kbest}). 
It explores the search tree level by level according to the BFS model. 
However, the algorithm keeps only the best \textit{K} nodes in terms of evaluation for 
further exploration, and the remaining nodes from the level are systematically removed.  
This process is repeated for each level until reaching the last one where leaf nodes (solutions) exist. 
Among these solutions, the algorithm returns the best one in terms of distance. 
Since the search tree of this algorithm contains $k$ nodes in each level, the total 
number of explored nodes by this algorithm is equal to $(M-1)\times K$, where $M$ refers 
to the number of transmit antennas. Thereby, this algorithm has a fixed complexity irrespective to SNR. 

Moreover, the number of kept nodes ($K$) should be carefully considered since it impacts the complexity of the algorithm.
On one hand, a large value of parameter $K$ allows the algorithm to achieve a near SD 
performance in terms of BER. However, the algorithm complexity increases
significantly and can even exceed the SD complexity. 
In addition to that, a large value of $K$ induces a significant sorting overhead,
making the complexity of K-best far from real-time response.     
On the other hand, a small value for parameter $K$ reduces the complexity, 
however, the algorithm loses in performance in terms of BER. 
Moreover, the performance of the algorithm, in terms of BER,
drops significantly for dense constellations. 
To overcome all these drawbacks, we propose a hybrid parallel algorithm,
named SD-K-best that takes all the benefits of SD and K-best algorithms.
Our hybrid approach also aims  to reduce the complexity of our high-level
parallel SD version, while taking benefit from the Best-FS exploration, the sphere radius,
and the diversification gain.

Indeed, our hybrid approach is based on our high-level parallel scheme, which
means that a  Master/Worker paradigm is also used in this approximate approach. 
The mater process executes a SD instance which builds the SD search tree in the master work-pool. 
To accelerate the exploration process of this tree, we use several  workers with a low-complexity K-best algorithm.  
In other words, the master performs the pre-processing phase and generates the root of the SD search tree.  
After-that, it explores the search tree by a SD algorithm according to the Best-FS 
exploration mode which allows to explore first the most promising combinations. 
To accelerate the exploration process, the master sends nodes, 
from the head of the master work-pool (right-most nodes),  for workers to complete their exploration. 
Each worker explores the received node by the mean of a K-best algorithm. 
Therefore, after the branching a worker keeps only the best $K$ successors in
terms of partial distance and ignore the rest which reduces the complexity. 
In-addition to the best $K$  nodes selected by each worker, 
additional nodes may also be selected if their distance is very close to the distance of selected nodes. 
Moreover, since the value of parameter $K$ used by each worker is small, there is an insignificant sorting overhead
 which allows workers to rapidly reach  leaf-nodes and improve the radius. 
Improving the radius allows to reduce even more the complexity of workers. i.e.,  
workers keep only the best successors which are inside the radius. This allows to target better quality combinations. 
In the case where  the kept nodes by a worker are outside the radius, 
the corresponding worker ends its exploration and requests a new node from the master.  
The fact that the search tree is built in parallel allows to take benefit from the diversification gain 
which may allow to explore good combination and thus reducing more efficiently the 
radius and avoiding the explorations of huge number of branches. 
Finally,  the ends of this hybrid approach is reached when the master work-pool is empty. 
The whole number of explored nodes by our hybrid SD-K-best is much bigger than the number
of explored nodes by the K-best algorithm which leads to improve the BER performance.
However, since this approach takes benefit of parallel architectures, 
the complexity of our hybrid SD-K-best can be less than the complexity of K-best algorithm, 
since the average number of explored nodes by a processing element is lower. 
In this way, we have been able to achieve both low complexity and good BER
performance as demonstrated in the experiment results section.

\section{Experimental Results}
\label{sec:results}
In this section, we report the results of our proposed approaches. This section is organized in three parts. 
The first part shows the impact of different exploration strategies and  
optimization techniques on the SD complexity.  
The second part of this section reports the impact of using parallel architectures
to accelerate the exploration process of the SD algorithm. Finally, the third part presents the 
results, in terms of BER performance and time complexity, of our proposed approximate 
algorithms for large MIMO systems.

We perform our experiments using a two-socket 10-core Intel Ivy Bridge CPU
running at 2.8 GHz with 256 GB of main memory. Hyper-threading is enabled on the system
in order to maximize resource occupancy.
For all the experiments, we consider the case of a perfect channel-state information. 
This means that the channel matrix is known only at the receiver. 

\subsection{Results of Optimizing the SD components}
In the following, we report the results of using different exploration strategies 
and optimization techniques on the SD complexity. 

\begin{figure}
 \centering
     \begin{subfigure}[b]{0.4\textwidth}
    \begin{tikzpicture}[scale=0.6,font=\small]
    \renewcommand{\axisdefaulttryminticks}{4}
    \tikzstyle{every major grid}+=[style=densely dashed]
    \tikzstyle{every axis y label}+=[yshift=-10pt]
    \tikzstyle{every axis x label}+=[yshift=5pt]
    \tikzstyle{every axis legend}+=[cells={anchor=west},fill=white,
        at={(0.4,0.97)}, anchor=north west, font=\small]
    \begin{semilogyaxis}[
      xmin=0,
			title={(a) Complexity results},
      ymin=3*10^-5,
      xmax=26,
      ymax=3*10^4,
      grid=major,
      scaled ticks=true,
    xlabel={SNR},
    ylabel={Decoding Time (s) } ]

   \addplot[color=blue,mark size=2pt,mark=circle*,line width=1pt] coordinates{ 
(0,1529)(4,1394)(8,134)(12,0.357)(16,0.010652)(20, 0.001)(24,0.0002)
  };                               \addlegendentry{SD with BFS}
  
    \addplot[color=red,mark size=3pt,mark=triangle*,line width=1pt] coordinates{ 
(0,0.0219)(4,0.01523 )(8,0.013)(12,0.0076307)(16,0.002632 )(20,0.000403)(24,0.000196 )
  };                               \addlegendentry{SD with DFS}

 \addplot[color=black,mark size=3pt,mark=square*,line width=1pt] coordinates{ 
(0,0.006464 )(4,0.002)(8,0.0006 )(12,0.0003405)(16,0.0002)(20,0.0002)(24,0.000182)
  };                               \addlegendentry{SD with Best-FS}
              \end{semilogyaxis}
  \end{tikzpicture} \vskip-4mm
\centering
  \end{subfigure}
  \begin{subfigure}[b]{0.4\textwidth}
	
  \centering
   \begin{tikzpicture}[scale=0.6,font=\small]
    \tikzstyle{every major grid}+=[style=densely dashed]
    \begin{semilogyaxis}[
      ymin=10^0,
			title={(b) Number of  visited search-tree nodes.},
      ymax=4*10^6,
       ybar,
      bar width=14pt,
      grid=major,
     scaled ticks=true,
    xlabel={SNR},
    ylabel={Visited Nodes},
    xtick=data,
      axis x line=bottom,
      axis y line=left,
     enlarge x limits=0.2,
      symbolic x coords={0,12,24},
       xticklabel style={anchor=base,yshift=-\baselineskip}
      ]

   \addplot[fill=blue] coordinates{ 
(0,263690)(12,31134)
(24,48) };  \addlegendentry{SD with BFS}
     \addplot[fill=red] coordinates{ 
(0,4380)(12,1988)
(24,49) };        \addlegendentry{SD with DFS}
   \addplot[fill=black] coordinates{ 
(0,1203)(12,99) (24,36) }; \addlegendentry{SD with Best-FS}
\end{semilogyaxis}
  \end{tikzpicture} \vskip-4mm
\centering
 	\end{subfigure}
  \caption{	Impact of exploration strategies on SD performance
for the $18 \times 18$ MIMO system using BPSK modulation.}
\label{fig:visitednodes}
\end{figure}
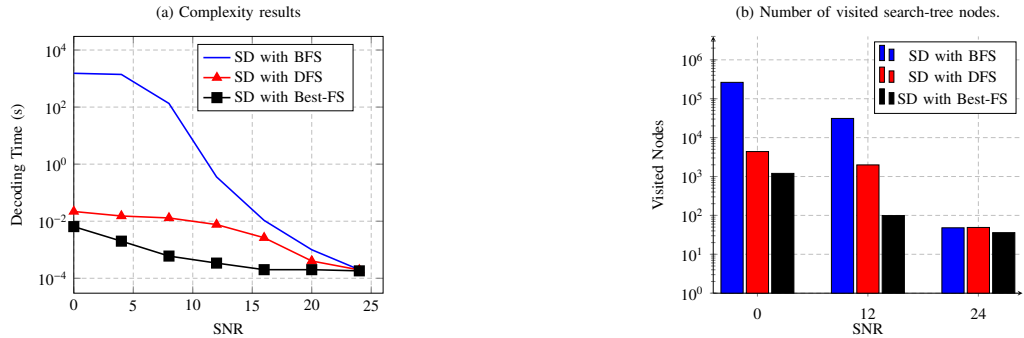

Figure \ref{fig:visitednodes} reports the 
complexity and the number of visited tree nodes by our SD algorithm using different
exploration strategies for a 18 $\times$ 18 MIMO system with BPSK modulation. 
The first result from Figure \ref{fig:visitednodes} is the positive impact of depth 
exploration strategies (DFS and Best-FS) on the SD algorithm complexity, 
as compared to the BFS exploration, especially in low SNR region. 
Indeed, the DFS is faster than BFS in low SNR region. 
This huge difference in complexity is the result of reducing the number
of explored nodes, as reflected in Figure \ref{fig:visitednodes} (b).
Indeed, the goal of the DFS is to reach leaf nodes quickly, which allows to explore a
large number of combinations (leaf nodes). Thus, reducing the sphere radius and avoiding 
the exploration of non-promising branches, which reduces the complexity. This is not the case of the BFS. 
This latter explores the search tree level by level until reaching the last one where solutions exist. 
Therefore, the sphere radius remains the same during the decoding process.  
In-addition, the majority of nodes explored by the BFS belongs to lower levels of the search tree 
where the evaluation process has higher complexity as compared to the early levels of the tree. 
To summarize, the BFS explores huge number of nodes with high evaluation complexity as compared to the DFS. 
The latter avoids non-promising branches at early levels of the search tree. 
According to the same figures, the Best-FS performs even better as
compared to the complexity of using the BFS and DFS strategies.  
Indeed, the Best-FS guides the search process toward better quality leaf nodes, which reduces 
the sphere radius more quickly and efficiently. Therefore, reducing further the number 
of explored nodes as compared to the DFS. 
This explains why the complexity of SD using the Best-FS is better than the complexity of 
the SD using the DFS. 
Figure \ref{fig:visitednodes} (a) also shows that the difference in complexity between the different exploration 
strategies decreases when increasing the SNR until reaching the same complexity for SNR equals to 24.
This behavior is related to the estimation of the initial sphere radius. Indeed,
the higher the SNR (lower noise) the better the estimation of the initial sphere radius.
In fact, when reaching 24 dB, the initial sphere radius becomes smaller which induces an efficient pruning
 process for the BFS and equal number of explored nodes as we can see in Figure \ref{fig:visitednodes} (b).
Moreover, Figure \ref{fig:visitednodes} (a) also shows that using the SD algorithm
with Best-FS allowed us to reach real-time response ($10^{-2}s$) 
from 0 dB SNR, while the SD algorithm with DFS and BFS needs a SNR of 10 dB and 16 dB respectively to 
ensure a real-time decoding process. Thus, a difference of 16 dB in power consumption.
 
As a conclusion, the Best-FS is more suitable for serial implementation of the SD
algorithm. For this reason, we will use it for all remaining experiments.

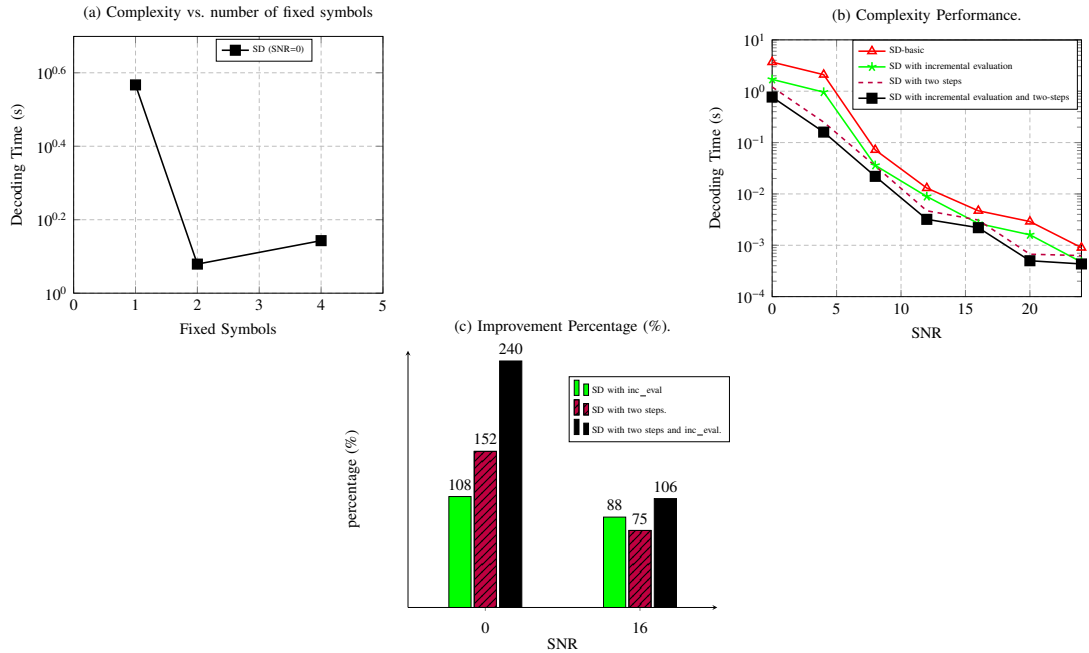
\begin{figure}
\begin{subfigure}[b]{0.5\textwidth}
  \centering
   \begin{tikzpicture}[scale=0.6,font=\small]
    \renewcommand{\axisdefaulttryminticks}{4}
    \tikzstyle{every major grid}+=[style=densely dashed]
    \tikzstyle{every axis y label}+=[yshift=-10pt]
    \tikzstyle{every axis x label}+=[yshift=5pt]
    \tikzstyle{every axis legend}+=[cells={anchor=west},fill=white,
        at={(0.46,0.99)}, anchor=north west, font=\tiny]
    \begin{semilogyaxis}[
      xmin=0,
				title={(a)  Complexity vs. number of fixed symbols},
      ymin=1,
      xmax=5,
      ymax=5,
      grid=major,
    xlabel={Fixed Symbols},
    ylabel={Decoding Time (s)} ]
\addplot[color=black,mark size=3pt,mark=square*,line width=1pt] coordinates{ 
(1,3.69 )(2,1.2)(4,1.39)
  };                               \addlegendentry{SD (SNR=0)} 

              \end{semilogyaxis}
  \end{tikzpicture} \vskip-4mm
\centering
  \label{fig:steps}
\end{subfigure}
	\hfill
 \begin{subfigure}[b]{0.48\textwidth}
  \centering
   \begin{tikzpicture}[scale=0.6,font=\small]
    \renewcommand{\axisdefaulttryminticks}{4}
    \tikzstyle{every major grid}+=[style=densely dashed]
    \tikzstyle{every axis y label}+=[yshift=-10pt]
    \tikzstyle{every axis x label}+=[yshift=5pt]
    \tikzstyle{every axis legend}+=[cells={anchor=west},fill=white,
        at={(0.26,1)}, anchor=north west, font=\tiny]
    \begin{semilogyaxis}[
      xmin=0,
			title={(b) Complexity Performance.},
      ymin=10^-4,
      xmax=24,
      ymax=10,
      grid=major,
      scaled ticks=true,
    xlabel={SNR},
    ylabel={Decoding Time (s) }]
                          
 \addplot[color=red,mark size=3pt,mark=triangle,line width=1pt] coordinates{ 
(0,3.69 )(4,2.10)(8,0.072 )(12,0.013)(16,0.0047)(20,0.0029)(24,0.0009)
  };                               \addlegendentry{SD-basic}
 
\addplot[color=green,mark size=3pt,mark=star,line width=1pt] coordinates{ 
(0,1.70 )(4,0.96)(8,0.036 )(12,0.0089)(16,0.00265)(20,0.0016)(24,0.00047)
  };                               \addlegendentry{SD with incremental evaluation}

  \addplot[color=purple, dashed ,mark size=3pt,mark=plus,line width=1pt]   coordinates{ 
(0,1.21)(4,0.25)(8,0.035)(12,0.0047)(16,0.0031)(20,0.00067)(24,0.00063)};
                                   \addlegendentry{SD with two steps } 

\addplot[color=black,mark size=3pt,mark=square*,line width=1pt]   coordinates{ 
(0,0.77)(4,0.16)(8,0.022)(12,0.0032)(16,0.0022)(20,0.0005)(24,0.000433)};
                                 \addlegendentry{SD with incremental evaluation and two-steps} 
  \end{semilogyaxis}
  \end{tikzpicture} \vskip-4mm
\centering
  \label{fig:inc}
	\end{subfigure}
	\hfill
 \begin{subfigure}[b]{1\textwidth}
  \centering
 \begin{tikzpicture}[scale=0.6,font=\small]
    \tikzstyle{every major grid}+=[style=densely dashed]
    \tikzstyle{every axis y label}+=[yshift=-10pt]
    \tikzstyle{every axis x label}+=[yshift=5pt]
    \tikzstyle{every axis legend}+=[cells={anchor=west},fill=white,
        at={(0.52,0.9)}, anchor=north west, font=\tiny ]
    \begin{axis}[
      ybar,
			title={(c)  Improvement Percentage (\%).},
      bar width=14pt,
      xlabel={SNR},
      ylabel={percentage (\%)},
      ymin=0,
			ymax=250,
      ytick=\empty,
      xtick=data,
      axis x line=bottom,
      axis y line=left,
      enlarge x limits=0.5,
      symbolic x coords={0,16},
      xticklabel style={anchor=base,yshift=-\baselineskip},
      nodes near coords={\pgfmathprintnumber\pgfplotspointmeta}
    ]
     \addplot[fill=green]coordinates{
        (0,108)
        (16,88)
              };\addlegendentry{SD with inc\_eval}

                \addplot[fill=purple, postaction={
        pattern=north east lines
    }] coordinates{
       (0,152)
       (16,75)
       };\addlegendentry{SD with two steps.}
                \addplot[fill=black] coordinates {
        (0,240)
        (16,106)
              };\addlegendentry{SD with two steps and inc\_eval. }
    \end{axis}
  \end{tikzpicture}
  \centering
  	\label{fig:imp}
		\end{subfigure}
		\hfill
		 \caption{ Impact of our optimization techniques on the SD performance 
  for a $52 \times 52 $ MIMO system using BPSK modulation.}
	\label{fig:opt}
\end{figure}

Figure \ref{fig:opt} shows the impact of our optimization techniques on the complexity
of SD algorithm for 52 $\times$ 52 MIMO system using BPSK modulation. 

The first sub-figure (a) reports the complexity of performing the branching over several
symbols at each iteration of the SD algorithm in low SNR region (0 dB). 
The subfigure shows that the best performance is reached when fixing two symbols
at a time. After that, increasing the number of fixed symbols increases the complexity. 
This is mostly due to the increase in the number of resulting nodes that need to be
evaluated; which induces an overhead in computation, 
especially when using a single CPU-core.
For this reason, it is important that the number of fixed symbols fits well the
targeted architecture and the number of its processing elements.

Similarly, sub-figures (b) and (c) report respectively the complexity over SNR and the improvement percentage of our
optimizations in two SNR regions. 
The sub-figures show the positive impact
of our incremental evaluation process on the complexity of the SD algorithm. 
The incremental evaluation process allowed us to increase the performance of the SD algorithm up to 108\%,
i.e., more than two times faster. Moreover, fixing two positions in 
the transmitted vector at a time, allowed an improvement in complexity up to 152\%. 
Combining both optimizations allowed us to achieve even higher
improvement to reach 240\%. In addition to the improvement in complexity, 
we also have an improvement of four dB in power consumption as compared to the basic SD version.   
This improvement is the results of (1) grouping some evaluation steps and (2)
avoiding redundancy in computing the evaluation for each search tree node.

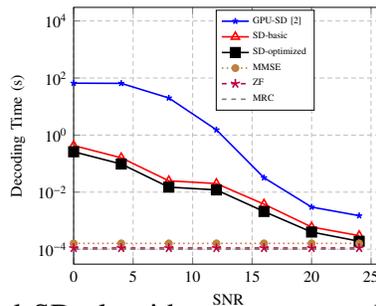
\begin{figure}
  \centering
   \begin{tikzpicture}[scale=0.6,font=\small]
    \renewcommand{\axisdefaulttryminticks}{4}
    \tikzstyle{every major grid}+=[style=densely dashed]
    \tikzstyle{every axis y label}+=[yshift=-10pt]
    \tikzstyle{every axis x label}+=[yshift=5pt]
    \tikzstyle{every axis legend}+=[cells={anchor=west},fill=white,
        at={(0.46,0.99)}, anchor=north west, font=\tiny]
    \begin{semilogyaxis}[
      xmin=0,
      ymin=3*10^-5,
      xmax=26,
      ymax=3*10^4,
      grid=major,
      scaled ticks=true,
    xlabel={SNR},
    ylabel={Decoding Time (s) } ]

   \addplot[color=blue,mark size=2pt,mark=star,line width=1pt] coordinates{ 
(0,66)(4,65)(8,20)(12,1.53)(16,0.032)(20,0.003)(24,0.0015)
  };                               \addlegendentry{GPU-SD \cite{arfaoui2016}}
  
                                         
 \addplot[color=red,mark size=3pt,mark=triangle,line width=1pt] coordinates{ 
(0,0.43)(4,0.16)(8,0.025)(12,0.02)(16,0.0038)(20,0.0006)(24,0.0003)
  };                               \addlegendentry{SD-basic}
  \addplot[color=black, mark size=3pt,mark=square*,line width=1pt]   coordinates{ 
(0,0.26)(4,0.097)(8,0.015)(12,0.012)(16,0.0021)(20,0.0004)(24,0.00019)};
                                   \addlegendentry{SD-optimized } 
   \addplot[color=brown,  dotted, mark=*, mark options={solid}, mark size=2pt,line width=1pt]   coordinates{ 
(0,0.00016)(4,0.00016)(8,0.00016)(12,0.00016)(16,0.00016)(20,0.00016)(24,0.00016)};
                                   \addlegendentry{MMSE} 
   \addplot[color=purple, dashed,  mark size=3pt,mark=star,line width=1pt]   coordinates{ 
(0,0.00011)(4,0.00011)(8,0.00011)(12,0.00011)(16,0.00011)(20,0.00011)(24,0.00011)};
                                   \addlegendentry{ZF} 
   \addplot[color=gray, dashed ,mark size=3pt,mark=Circle,line width=1pt]   coordinates{ 
(0,0.00010)(4,0.00010)(8,0.00010)(12,0.00010)(16,0.00010)(20,0.00010)(24,0.00010)};
                                   \addlegendentry{MRC}                                
   
              \end{semilogyaxis}
  \end{tikzpicture} \vskip-4mm
\centering
\caption{Complexity of our optimized SD algorithm as compared to our previous GPU-SD work
in \cite{arfaoui2016} for the $25 \times 25$ MIMO system.}
  \label{fig4a}
\end{figure}

Figure \ref{fig4a}  shows the complexity  of our approach as
compared to our previous work in \cite{arfaoui2016}
using a 25 $\times$ 25 MIMO system with BPSK modulation. The figure also
shows the complexity results of the most used
linear decoders in the literature: MMSE, ZF, and MRC. 
The  $25\times25$ MIMO system represents the biggest that can be solved by the SD 
algorithm with BFS, due to the huge memory resources required by this strategy.

Figure \ref{fig4a} shows the importance of carefully optimizing the SD
algorithm before moving toward its parallelization. 
Indeed, our optimized SD implementation is up to 255 times faster in the low
SNR region and 15 times faster in the high SNR region as compared to the GPU-based SD
in \cite{arfaoui2016}. 
Figure \ref{fig4a} shows that, unlike linear decoders which have constant complexity, 
the complexity of the SD algorithm decreases when increasing the SNR. 
This behavior is closely related to the fact that the initial
value of the sphere's radius, which is inversely proportional to the SNR. 
Therefore, higher the SNR, the smaller initial sphere radius,
which reduces the search space, and thus, the complexity.

\subsection{Results of our Parallel SD Approaches}
In the following, we report the impact of using parallel architectures on the SD complexity. 
Two parallel approaches have been proposed. The first parallel approach (PL-SD) uses a set of 
threads to explore several nodes at a time; while the second parallel approach (PSD) 
uses simultaneously several instances of the SD algorithm to explore the search tree in parallel. 
Due to the unbalanced work-load for each instance, two version of the PSD are proposed depending 
on the nature of the used load-balancing strategy: PSD with static load-balancing 
(S-PSD) and PSD with dynamic load-balancing (D-PSD). 
In our PL-SD approach, creating and destroying threads at each iteration
induces a considerable overhead time, which 
slows down considerably this parallel version. To overcome this problem, 
the workload for each parallel thread must be high enough to cover this overhead time. 
In our case, the workload for each thread is around twenty nodes. 

\begin{figure}
  \centering
	\begin{subfigure}[b]{0.4\textwidth}
   \begin{tikzpicture}[scale=0.6,font=\small]
    \tikzstyle{every major grid}+=[style=densely dashed]
    \tikzstyle{every axis y label}+=[yshift=-10pt]
    \tikzstyle{every axis x label}+=[yshift=5pt]
    \tikzstyle{every axis legend}+=[cells={anchor=west},fill=white,
        at={(0.3,0.85)}, anchor=north west, font=\tiny]
    \begin{axis}[
      xmin=2,
      ymin=0,
      xmax=40,
      ymax=1,
      grid=major,
    xlabel={Number of threads},
    ylabel={Decoding Time (s)} ]
	
		\addplot[color=black,mark size=3pt,mark=triangle,line width=1pt] coordinates{ 
(2,0.97 )(5,0.97)(10,0.97 )(15,0.97)(20,0.97)(25,0.97)
 (30,0.97)(35,0.97) (40,0.97)
  };                               \addlegendentry{Serial SD } 
	\addplot[color= blue,dashed,mark size=3pt,mark=star,line width=1pt] coordinates{ 
(2,0.3 )(5,0.154)(10,0.07 )(15,0.28)(20,0.15)(25,0.15)
 (30,0.17)(35,0.16) (40,0.15)
  };                               \addlegendentry{S-PSD} 
  
\addplot[color=green,mark size=3pt,mark=square*,line width=1pt] coordinates{ 
(2,0.32 )(5,0.16)(10,0.087 )(15,0.088)(20,0.061)(25,0.057)
 (30,0.058)(35,0.051) (40,0.089)
  };                               \addlegendentry{PL-SD} 

\addplot[color=blue,mark size=3pt,mark=star,line width=1pt] coordinates{ 
(2,0.28 )(5,0.07)(10,0.044 )(15,0.031)(20,0.026)(25,0.026)
 (30,0.023)(35,0.022) (40,0.023)
  };                               \addlegendentry{D-PSD}             
			\end{axis}
  \end{tikzpicture} \vskip-4mm
 \end{subfigure}
\begin{subfigure}[b]{0.4\textwidth}
  \begin{tikzpicture}[scale=0.48]
  \centering
  \begin{axis}[
        ybar, axis on top,
        title={s},
        height=8cm, width=15.5cm,
        bar width=0.32cm,
        ymajorgrids, tick align=inside,
        major grid style={draw=white},
        enlarge y limits={value=.1,upper},
        ymin=0, ymax=45,
        axis x line*=bottom,
        axis y line*=right,
        y axis line style={opacity=0},
        tickwidth=0pt,
        enlarge x limits=true,
        legend style={
            at={(0.5,-0.2)},
            anchor=north,
            legend columns=-1,
            /tikz/every even column/.append style={column sep=0.2cm}
        },
        ylabel={Speedup},
        symbolic x coords  ={
          2,5,10,15,20,
           25,30,35,40         },
       xtick=data,
       nodes near coords={
        \pgfmathprintnumber[precision=0]{\pgfplotspointmeta}
       }
    ]
    \addplot [draw=none, fill=blue,    postaction={
        pattern=north east lines
    }] coordinates {
      (2,4)
	  (5,7)
      (10, 5) 
      (15,16)
      (20,6.46) 
      (25,6.55) 
      (30,6.46)
			  (35,6.46)
      (40,6.46) 
  };
   \addplot [draw=none,fill=green] coordinates {
      (2,3.3)
	  (5,7)
      (10, 12) 
      (15,12)
      (20,18) 
      (25,21) 
      (30,20)
			  (35,21)
      (40,12) 
  };
   \addplot [draw=none, fill=blue] coordinates {
      (2,4.66)
	  (5,10)
      (10,20) 
      (15,26)
      (20,31) 
      (25,32) 
      (30,35)
			  (35,41)
      (40,40.69) 
	  };

    \legend{S-PSD ,PL-SD,D-PSD }
  \end{axis}
  \end{tikzpicture}
	\end{subfigure}
 \centering
\caption{Impact of increasing the number of threads on the complexity of
our parallel SD approaches for a $10 \times 10$ MIMO system using 16-QAM  modulation.} 
  \label{fig:low1}
\end{figure}
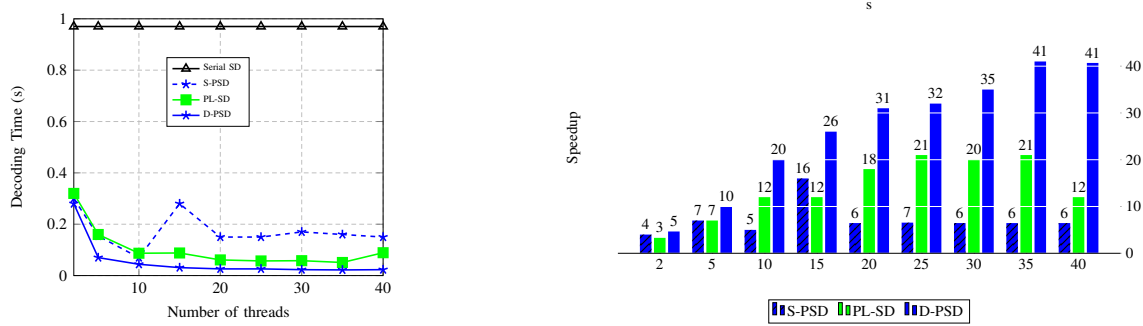

Figure\ref{fig:low1} shows the impact of increasing the number of
parallel threads on the time complexity of our parallel SD approaches. 
 This time is measured for SNR equals to zero.
Figure\ref{fig:low1} also shows the speedup obtained by
our approaches for each number of threads. 
The first result from the figure is the positive impact of parallelism
on reducing the complexity of the SD algorithm.  
As we can see in Figure \ref{fig:low1}, the curve of our
low-level parallel approach has three phases.
 A first phase, between two and ten threads, characterized by
 a rapid decrease in complexity, when increasing the number of threads. 
This means that adding threads in this phase is beneficial and reduces the complexity. 
This is due to the positive impact of splitting the work-load over
several processing elements (CPU-cores) and the low synchronization overhead. 
This overhead is related to the concurrent access to the same
data structure that contains the search tree. 
This latter must be accessed by one thread at a time to have a valid execution result. 
After that, begins a second phase between ten and thirty-five
threads where adding new threads has no impact of the decoding time. 
This can be explained by fact that the synchronization overhead
neutralizes the gain of exploiting additional processing elements. 
After that, a third phase begins. This last is characterized by an increase in complexity
when increasing the number of parallel threads. 
This behavior is related to the overhead of synchronization
which increases when increasing the number of threads. 
Along with the sequential execution of threads by processing elements.   
As a result, an ideal number of threads for this low-level version
will be between twenty and thirty for this parallel approach. 

Figure \ref{fig:low1} also shows the results of our high-level
parallel approach with static and dynamic load-balancing strategies.
This parallel version is based on splitting the search tree
into several subtrees, where each one is explored independently by a parallel thread.
The gain obtained by our high-level approach using a static load-balancing (S-PSD) strategy   
is limited due to the imbalanced work-load in each subtree. 
This results in a long execution time for few threads while others are ideal.  
By adding our dynamic load-balancing strategy (D-PSD), the performance
of our high level parallel approach improved substantially. 
In fact, adding our proposed dynamic load-balancing strategy allowed us to reach
a relative speedup of forty times faster using twenty 
CPU-cores and thirty-five parallel threads as compared to our optimized serial SD version. 
The curve of our high-level approach has two phases. 
A first phase characterized by a rapid decrease in complexity
(increase in speedup), and a second phase where adding new threads
does not change the complexity.
The first phase, between two and twenty threads has a super-linear speedup. 
This speedup is the result of (1) low synchronization 
overhead since each thread explores its subtree independently from the others. 
Therefore, no concurrent access to the same data structure. 
(2) The fair work-load distribution among parallel threads due 
to our load-balancing strategy. This latter prevents the idleness of threads.  
(3) The diversification gain, that allows to reduce 
the explored search space as compared to the serial version.  
Indeed, dividing  and exploring the search tree in parallel may result in a
rapid improvement of the radius, which allows to avoid the 
exploration of several branches explored in the sequential version.
This results in a super-linear speedup as in our case. 
Moreover, when the number of threads is greater than twenty which is the number
of available processing elements (CPU-cores), a second phase begins. 
In this phase,  the complexity of our high-level approach should be
increasing due to the serial execution of threads by the processing elements. 
However, this is not the case. In fact, we still get performance
improvement to reach 41x speedup when using forty threads.  
This behavior can simply be explained by the diversification
gain that neutralizes the serial execution overhead of threads and improve the performance. 

\begin{figure}
  \centering
	\begin{subfigure}[b]{0.3\textwidth}
   \begin{tikzpicture}[scale=0.5,font=\small]
    \renewcommand{\axisdefaulttryminticks}{4}
    \tikzstyle{every major grid}+=[style=densely dashed]
    \tikzstyle{every axis y label}+=[yshift=-10pt]
    \tikzstyle{every axis x label}+=[yshift=5pt]
    \tikzstyle{every axis legend}+=[cells={anchor=west},fill=white,
        at={(0.6,1.06)}, anchor=north west, font=\tiny ]
    \begin{semilogyaxis}[
      xmin=0, axis on top,
  title={(a) Complexity results of our parallel SD approaches.},
      ymin=10^-3,
      xmax=26,
      ymax=10^2,
      grid=major,
      scaled ticks=true,
    xlabel={SNR},
    ylabel={Decoding Time (s) } ]



\addplot[color=red,mark size=2pt,line width=1pt] coordinates{ 
(0,127.33)(4,23)(8,1.62)(12,1.47)(16,0.24)(20,0.062)(24,0.002)
  };                               \addlegendentry{Serial SD $16\times16$  }
  
 \addplot[color=red,dashed,mark size=2pt,mark=square,line width=1pt] coordinates{ 
(0,7.86)(4,1.86)(8,1.32)(12,0.629)(16,0.371)(20,0.3)(24,0.3)
  };                               \addlegendentry{PL-SD $16\times16$  } 

  \addplot[color=red,dashed,mark size=2pt,mark=star,line width=1pt] coordinates{ 
(0,2.1)(4,0.29)(8,0.14)(12,0.06)(16,0.0068)(20,0.0038)(24,0.003)
  };                               \addlegendentry{ D-PSD $16\times16$ }

  \addplot[color=black,mark size=2pt,line width=1pt] coordinates{ 
(0,0.9731)(4,0.44)(8,0.1)(12,0.0586)(16,0.02)(20,0.0083)(24,0.0022)
  };                               \addlegendentry{Serial SD $10\times10$  }
\addplot[color=black,dashed,mark size=2pt,mark=square,line width=1pt] coordinates{ 
(0,0.07)(4,0.03)(8,0.015)(12,0.013)(16,0.01)(20,0.01)(24,0.009)
  };                               \addlegendentry{Pl-SD $10\times10$  }  

\addplot[color=black,dashed,mark size=2pt,mark=star,line width=1pt] coordinates{ 
(0,0.022)(4,0.01)(8,0.007)(12,0.005)(16,0.0022)(20,0.0019)(24,0.0019)
  };                               \addlegendentry{D-PSD $10\times10$  }
  
       \end{semilogyaxis}
  \end{tikzpicture} \vskip-5mm \hskip5mm 
  \end{subfigure}
  \vspace{1cm}
  \begin{subfigure}[b]{0.3\textwidth}
\begin{tikzpicture}[scale=0.4]
  \centering
  \begin{axis}[
        ybar, axis on top,
        title={(b) Speedup of our D-PSD  parallel approach.},
        height=8cm, width=15.5cm,
        bar width=0.32cm,
        ymajorgrids, tick align=inside,
        major grid style={draw=white},
        enlarge y limits={value=.1,upper},
        ymin=0, ymax=70,
        axis x line*=bottom,
        axis y line*=right,
        y axis line style={opacity=0},
        tickwidth=0pt,
        enlarge x limits=true,
        legend style={
            at={(0.5,-0.2)},
            anchor=north,
            legend columns=-1,
            /tikz/every even column/.append style={column sep=0.2cm}
        },
        ylabel={Speedup},
        symbolic x coords  ={
           0,4,8,12,16,
           20,24
           },
       xtick=data,
       nodes near coords={
        \pgfmathprintnumber[precision=0]{\pgfplotspointmeta}
       }
    ]
     \addplot [draw=none,fill=red] coordinates {
 (0,61)        
 (4,69)
      (8, 11.5) 
      (12,24.5)
      (16,24) 
(20,16)  
      (24,1)
      };
     \addplot [draw=none, fill=black] coordinates {
 (0,41)     
(4,44)
      (8, 14) 
      (12,11)
      (16,9) 
      (20,4) 
      (24,1.15)
      };

    \legend{$16 \times 16$,$10 \times 10$}
  \end{axis}
  \end{tikzpicture}
	\end{subfigure}
		\begin{subfigure}[b]{0.3\textwidth}
	 \centering
  \begin{tikzpicture}[scale=0.5,font=\small]
    \renewcommand{\axisdefaulttryminticks}{4}
    \tikzstyle{every major grid}+=[style=densely dashed]
    \tikzstyle{every axis y label}+=[yshift=-10pt]
    \tikzstyle{every axis x label}+=[yshift=5pt]
    \tikzstyle{every axis legend}+=[cells={anchor=west},fill=white,
        at={(0.55,0.98)}, anchor=north west, font=\small ]
    \begin{semilogyaxis}[
      xmin=10,
			title={(c) Error rate of our PSD algorithm.},
      ymin=10^-6,
      xmax=17,
      ymax=*10^0,
      grid=major,
      scaled ticks=true,
      xlabel={SNR},
    ylabel={Symbol Error Rate} ]
     \addplot[color=black,  mark size=3pt,mark=square,line width=1pt] coordinates{ 
(10,0.26)(12,0.06)(14,0.0056)(16,0.00012)(17,0.000012)
  };                               \addlegendentry{PSD $10 \times 10$}  
       \addplot[color=red,  mark size=3pt,mark=star,line width=1pt] coordinates{ 
(10,0.034)(12,0.0038)(13,0.00052)(14,0.0000216)(15,0.00000108)(16,0)
  };                               \addlegendentry{PSD $16 \times 16$}  
  \end{semilogyaxis}
  \end{tikzpicture} \vskip1mm
\centering
\end{subfigure}
  \centering
  \caption{Complexity of parallel and sequential SD algorithms with
  16-QAM modulation for $10 \times 10$ and $16 \times 16$ MIMO systems.}
  \label{fig9a}
\end{figure}
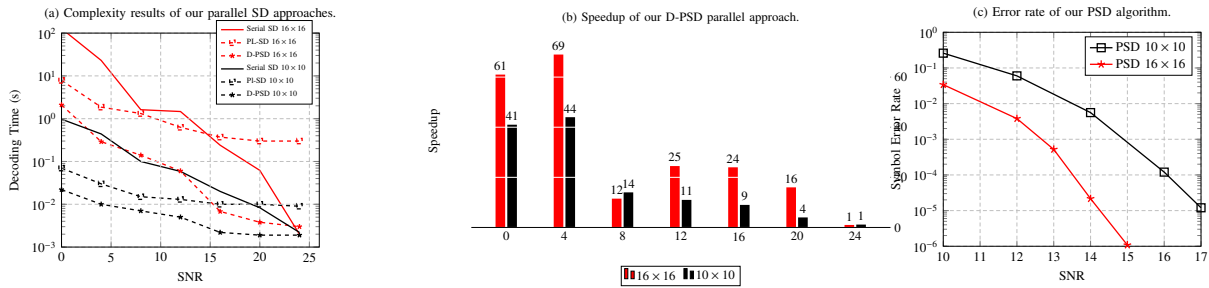
The scalability refers to the possibility of still improving the performance
of our parallel approaches when using huge number of processing elements. 
According to Figure \ref{fig:low1}, our low-level approach does not scale well since all parallel
threads operates on a single search tree. Therefore, inducing a synchronization 
overhead which effects the performance. This is not the case of our high-level 
approach which is embarrassingly parallel due to the low communication
and synchronization costs between parallel threads, 
since each one operates on its own search tree. However, it needs a
load balancing strategy to ensure a fair workload distribution among parallel threads.

\begin{figure}
  \centering
		\begin{subfigure}[b]{0.4\textwidth}
   \begin{tikzpicture}[scale=0.6,font=\small]
    \renewcommand{\axisdefaulttryminticks}{4}
    \tikzstyle{every major grid}+=[style=densely dashed]
    \tikzstyle{every axis y label}+=[yshift=-10pt]
    \tikzstyle{every axis x label}+=[yshift=5pt]
    \tikzstyle{every axis legend}+=[cells={anchor=west},fill=white,
        at={(1.01,1)}, anchor=north west, font=\small ]
    \begin{semilogyaxis}[
      xmin=10, axis on top,
	   title={(a) Latency in terms of visited nodes.},
      ymin=5,
      xmax=17,
      ymax=4*10^4,
      grid=major,
      scaled ticks=true,
    xlabel={SNR},
    ylabel={visited nodes (s) } ]

	\addplot[color=blue,mark size=2pt,mark=star,line width=1pt] coordinates{ 
(10,4194)(12,3602)(14,1797)(16,1505)
  };                               \addlegendentry{serial-SD }

\addplot[color=red,mark size=2pt,mark=star,line width=1pt] coordinates{ 
(10,2231)(12,1341)(14,717)(16,229)
  };                               \addlegendentry{S-PSD-16  }
  
	\addplot[color=green,mark size=2pt,,mark=star,line width=1pt] coordinates{ 
(10,2182)(12,1296)(14,695)(16,212)
  };                               \addlegendentry{S-PSD-32  }
  	\addplot[color=green,dashed,mark size=2pt,line width=1pt] coordinates{ 
(10,1020)(12,402)(14,140)(16,61)
  };                               \addlegendentry{MultiSphere-16 }
	
  \addplot[color=red,dashed,mark size=2pt,line width=1pt] coordinates{ 
(10,576)(12,207)(14,65)(16,28)
  };                               \addlegendentry{MultiSphere-32 }
  
	\addplot[color=green,mark size=2pt,line width=1pt] coordinates{ 
(10,139)(12,97)(14,40)(16,18)
  };                               \addlegendentry{D-PSD-16  }
	\addplot[color=red,mark size=2pt,line width=1pt] coordinates{ 
(10,69)(12,49)(14,21)(16,10)
  };                               \addlegendentry{D-PSD-32  }
		\end{semilogyaxis}
  \end{tikzpicture} 
  \centering
	 	\end{subfigure}
 	\begin{subfigure}[b]{0.4\textwidth}
 	\begin{tikzpicture}[scale=0.6,font=\small]
    \renewcommand{\axisdefaulttryminticks}{4}
    \tikzstyle{every major grid}+=[style=densely dashed]
    \tikzstyle{every axis y label}+=[yshift=-10pt]
    \tikzstyle{every axis x label}+=[yshift=5pt]
    \tikzstyle{every axis legend}+=[cells={anchor=west},fill=white,
        at={(1.01,1)}, anchor=north west, font=\small ]
	\begin{semilogyaxis}[
      xmin=10, axis on top,
	   title={(b) Complexity in terms evaluation (Partial Distance (PD)).},
      ymin=0,
      xmax=17,
      ymax=300000,
      grid=major,
      scaled ticks=true,
    xlabel={SNR},
    ylabel={PD calculations (s) } ]
	
	
 		\addplot[color=blue,mark size=2pt,mark=star,line width=1pt] coordinates{ 
(10,65452)(12,55932)(14,26957)(16,22150)
  };                               \addlegendentry{serial-SD }
  
  \addplot[color=red,mark size=2pt,mark=star,line width=1pt] coordinates{ 
(10,34712)(12,20804)(14,10575)(16,2744)
  };                               \addlegendentry{S-PSD-16  }
  
	\addplot[color=green,mark size=2pt,mark=star*,line width=1pt] coordinates{ 
(10,34316)(12,20190)(14,10440)(16,2682)
  };                               \addlegendentry{S-PSD-32  }
	
\addplot[color=red,dashed,mark size=2pt,line width=1pt] coordinates{ 
(10,5508)(12,2327)(14,1393)(16,1160)
  };                               \addlegendentry{MultiSphere-32  }
	\addplot[color=green,dashed,mark size=2pt,line width=1pt] coordinates{ 
(10,7179)(12,3136)(14,1122)(16,890)
  };                               \addlegendentry{MultiSphere-16  }
	\addplot[color=black,dashed,mark size=2pt,line width=1pt] coordinates{ 
(10,5483)(12,2936)(14,1727)(16,1033)
  };                               \addlegendentry{Geosphere}
		\addplot[color=green,mark size=2pt,line width=1pt] coordinates{ 
(10,2234)(12,1563)(14,656)(16,308)
  };                               \addlegendentry{D-PSD-16  }
 \addplot[color=red,mark size=2pt,line width=1pt] coordinates{ 
(10,1118)(12,801)(14,352)(16,179)
  };                               \addlegendentry{D-PSD-32  }

	       \end{semilogyaxis}
  \end{tikzpicture} \vskip-5mm
	 	\end{subfigure}
  \caption{Our PSD's latency (a), and  complexity (b)
vs. our sequential SD and MultiSphere parallel approach for a $10 \times 10$ 16-QAM
MIMO.}
  \label{fig10e}
  \end{figure}
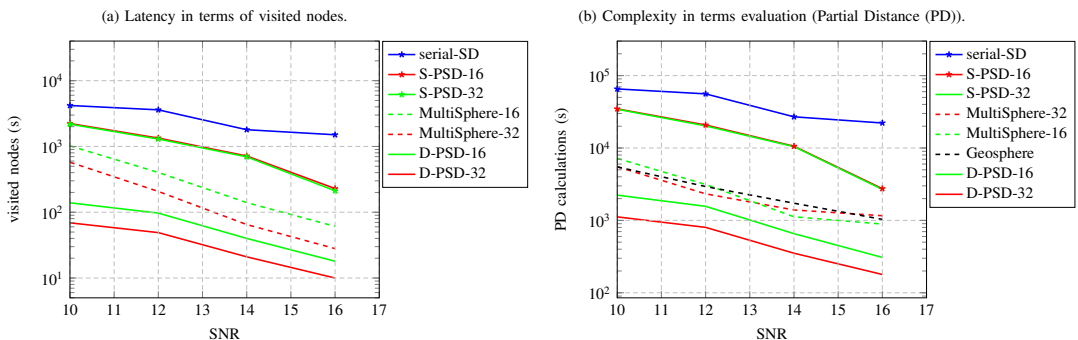

Figure \ref{fig9a} shows the performance (complexity, speedup) of our parallel
approaches when increasing the SNR for two MIMO systems ($10 \times 10$ and $
16 \times 16 $ ) using 16-QAM modulation. Our both parallel approaches use thirty-five
threads. 
Moreover, the speedup shown in this figure is obtained by our high-level parallel approach (D-PSD). 
The first observation is that the performance of our parallel approaches decreases when increasing the SNR.
This is caused by the decrease in work-load for each parallel process when increasing
the SNR since the sphere radius get smaller.
Our high-level approach gives better performance, for these
two MIMO configurations, as compared to our low level approach.  
This latter, does not perform well for high SNR and has higher 
complexity as compared to the sequential SD due to the synchronization overhead.
The figure shows a good performance of our high-level parallelization scheme,  
especially for SNR between 0 and 8 db, where we have been able to reach a speedup 
around 69 times faster as compared to our optimized sequential SD algorithm. 
This speedup is obtained by using 35 threads and exploiting 20 processing elements (CPU-cores).
This speedup is the result of: (1) reducing the synchronization overhead by choosing a good parallelization scheme, 
(2) exploiting efficiently the available computing resources by
dividing fairly the work-load using load balancing strategy, and 
(3) taking advantage of the diversification gain which is not well studied in the literature.    
Moreover, Figure \ref{fig9a}c shows the optimal Symbol Error Rate of our
PSD for a $10 \times 10$ and $16 \times 16$  MIMO systems using 16-QAM modulation.

In the following, we compare our high-level parallel approaches against
most recent works in the literature  \cite{nikitopoulos2018massively, nikitopoulos2014geosphere}. 
Figure \ref{fig10e}  shows a comparison between the latency and complexity of our PSD for two cases (16 and 32 threads) 
with the state-of-the-art parallel MultiSphere \cite{nikitopoulos2018massively}, 
Geosphere SDs \cite{nikitopoulos2014geosphere}, and our serial SD implementation
for 16-QAM modulation using a $10 \times 10$ MIMO system. 
For fair comparison, the initial sphere radius is set to infinite. 
The number \textit{visited nodes} refers to the average number of nodes (per thread)
on which we performed a branching process, 
and the \textit{PD calculations} refers to the average number of partial distance calculations, 
i.e., the average number of evaluated search tree nodes per thread. 
The PD calculations represents an important  factor that highly influences the complexity of decoding approaches.  
For our S-PSD, these numbers are measured for the thread with the largest work-load, 
since the complexity of the parallel version depends depends on the complexity of this thread. 
 
Figure \ref{fig10e} validates the results of our proposed parallel approach (PSD),
especially when using dynamic load balancing strategy. 
For both 16 and 32 cases, the results of D-PSD outperform the results of parallel 
MultiSphere\cite{nikitopoulos2018massively} and serial Geosphere \cite{nikitopoulos2014geosphere}
in terms of both visited nodes and PD calculations. 
Indeed, when increasing the number of threads from 16 to 32, our D-PSD consistently 
improves the performance in terms of both visited nodes and PD calculations. 
This is not the case of the MultiSphere approach and our S-PSD and in terms of 
PD calculations due to the unbalanced workload between parallel threads.   
Moreover, our D-PSD reduces the complexity by a factor of 58x as compared to our
serial SD and 5x as compared to the MultiSphere parallel approach.


\begin{figure}[t ]
  \centering
   \begin{tikzpicture}[scale=0.65,font=\small]
    \renewcommand{\axisdefaulttryminticks}{4}
    \tikzstyle{every major grid}+=[style=densely dashed]
    \tikzstyle{every axis y label}+=[yshift=-10pt]
    \tikzstyle{every axis x label}+=[yshift=5pt]
    \tikzstyle{every axis legend}+=[cells={anchor=west},fill=white,
        at={(0.5,0.98)}, anchor=north west, font=\small ]
    \begin{semilogyaxis}[
      xmin=0, axis on top,
	   title={(a) Complexity results of our SD-K-best approach.},
      ymin=10^-4,
      xmax=24,
      ymax=4*10^1,
      grid=major,
      scaled ticks=true,
    xlabel={SNR},
    ylabel={Decoding Time (s) } ]

\addplot[color=red,mark size=2pt,line width=1pt] coordinates{ 
(0,38.33)(4,3.72)(8,0.8)(12,0.29)(16,0.06)(18,0.06)(20,0.034)(24,0.0055)
  };                               \addlegendentry{D-PSD-30 }
  
	  \addplot[color=red,dashed,mark size=2pt,mark=square,line width=1pt] coordinates{ 
(0,0.025)(4,0.0226)(8,0.018)(12,0.014)(14,0.009)(16,0.006)(18,0.004)(20,0.003)(24,0.0026)
  };                               \addlegendentry{SD-K-best} 

       \end{semilogyaxis}
  \end{tikzpicture} \vskip-5mm
   \caption{Time complexity  of our SD-K-best algorithm vs. D-PSD algorithm
   for a $18 \times 18$ MIMO system using 16-QAM modulation.}
  \label{fig:sd-k-best01}
  \end{figure}
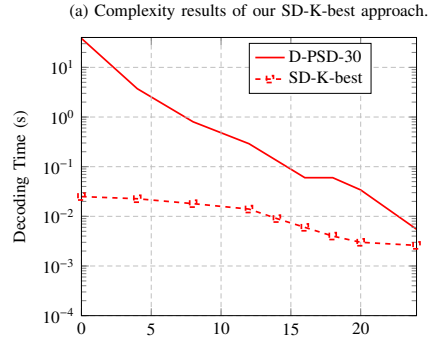

\subsection{Results of Approximate Algorithms}
The parallelization allows to improve the performance of
the SD algorithm especially, for large MIMO systems. 
However, the complexity remains very high to meet real-time response requirements for massive MIMO systems. 
To deal with such MIMO instances, the use of approximate algorithms is unavoidable. 
In the following, we report the preliminary results, as a proof of concept,
of our proposed approximate algorithm SD-K-best and compare its
performance against the well known K-best algorithm.   
The main idea of this hybrid algorithm is to accelerate the exploration
of the SD search tree, stored in the master process, 
in a approximate way by using several low-complexity K-best algorithms
(workers) running in parallel to meet real-time requirement.

Figure \ref{fig:sd-k-best01} shows a comparison between the performance, in terms of
complexity, of our SD-K-best approach as compared 
to our Dynamic PSD approach for a $18 \times 18$ MIMO system with 16-QAM modulation.  
We can see from the figure that our SD-K-best performs better in terms of complexity as compared to our PSD since it explores partially the search tree. 
This makes it more applicable for various cases. 
Indeed,  our SD-K-best approach is able to meet real time response from 13 dB, while our 
parallel PSD reaches this complexity at 23 dB; thus, 10 dB difference in power consumption.

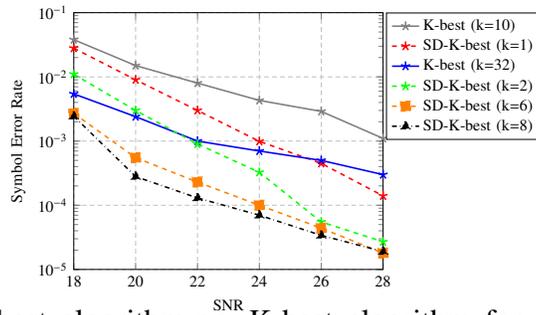
\begin{figure}
\centering
  \begin{tikzpicture}[scale=0.6,font=\small]
    \renewcommand{\axisdefaulttryminticks}{4}
    \tikzstyle{every major grid}+=[style=densely dashed]
    \tikzstyle{every axis y label}+=[yshift=-10pt]
    \tikzstyle{every axis x label}+=[yshift=5pt]
    \tikzstyle{every axis legend}+=[cells={anchor=west},fill=white,
        at={(1.01,1)}, anchor=north west, font=\small ]
    \begin{semilogyaxis}[
      xmin=18,
      ymin=10^-5,
      xmax=28,
      ymax=*10^-1,
      grid=major,
      scaled ticks=true,
      xlabel={SNR},
    ylabel={Symbol Error Rate} ]
     \addplot[color=gray,  mark size=3pt,mark=star,line width=1pt] coordinates{
(18,0.038)(20,0.015)(22,0.008)(24,0.0043)(26,0.0029)(28,0.0011)
  };                               \addlegendentry{K-best (k=10)}  
 \addplot[color=red, dashed,  mark size=3pt,mark=star,line width=1pt] coordinates{
(18,0.028)(20,0.009)(22,0.003)(24,0.00099)(26,0.00045)(28,0.00014)
  };                               \addlegendentry{SD-K-best (k=1)}
	    \addplot[color=blue,  mark size=3pt,mark=star,line width=1pt] coordinates{
(18,0.005435)(20,0.0024)(22,0.001)(24,0.0007)(26,0.0005)(28,0.0003)
  };                               \addlegendentry{K-best (k=32)}  

  \addplot[color=green , dashed,  mark size=3pt,mark=star,line width=1pt]
coordinates{
(18,0.011)(20,0.003)(22,0.0009)(24,0.000326)(26,0.000055)(28,0.000027)
  };                               \addlegendentry{SD-K-best (k=2)}  
  \addplot[color=orange, dashed,  mark size=3pt,mark=square*,line width=1pt] coordinates{
(18,0.0027)(20,0.0005513)(22,0.00023)(24,0.0001)(26,0.000044)(28,0.000018)
  };                               \addlegendentry{SD-K-best (k=6)}  
  \addplot[color=black, dashdotted,  mark size=3pt,mark=triangle*,line width=1pt] coordinates{
(18,0.00244)(20,0.00028)(22,0.00013)(24,0.00007)(26,0.000034)(28,0.000019)
  };                               \addlegendentry{SD-K-best (k=8)}

  \end{semilogyaxis}
  \end{tikzpicture} \vskip-4mm
\centering
  \caption{SER of our SD-K-best algorithm vs. K-best algorithm
  for a $16 \times 16$ MIMO system using 64-QAM modulation.}
  \label{fig:sd-k-best}
\end{figure}

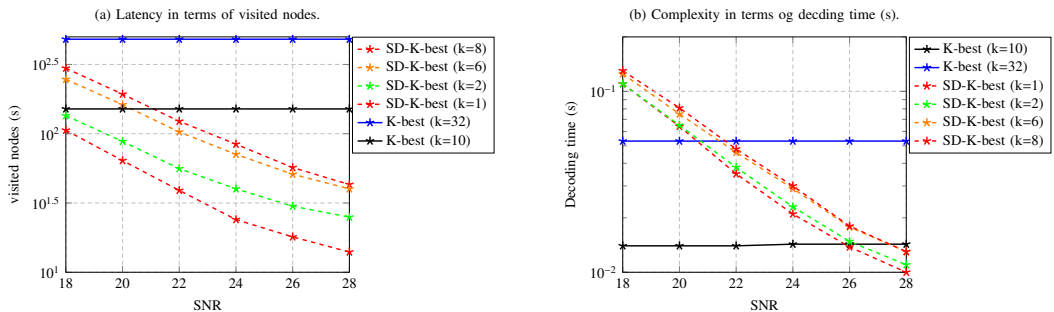
\begin{figure}
  \centering
\begin{subfigure}[b]{0.4\textwidth}
  \centering
   \begin{tikzpicture}[scale=0.55,font=\small]
    \renewcommand{\axisdefaulttryminticks}{4}
    \tikzstyle{every major grid}+=[style=densely dashed]
    \tikzstyle{every axis y label}+=[yshift=-10pt]
    \tikzstyle{every axis x label}+=[yshift=5pt]
    \tikzstyle{every axis legend}+=[cells={anchor=west},fill=white,
        at={(1.01,1)}, anchor=north west, font=\small ]
    \begin{semilogyaxis}[
      xmin=18, axis on top,
  title={(a) Latency in terms of visited nodes.},
      ymin=10,
      xmax=28,
      ymax=500,
      grid=major,
      scaled ticks=true,
    xlabel={SNR},
    ylabel={visited nodes (s) } ]
    \addplot[color=red, dashed,  mark size=3pt,mark=star,line width=1pt] coordinates{
(18,297)(20,193)(22,123)(24,84)(26,57)(28,43)
  };                               \addlegendentry{SD-K-best (k=8)}  
    \addplot[color=orange, dashed,  mark size=3pt,mark=star,line width=1pt] coordinates{
(18,247)(20,162)(22,103)(24,71)(26,51)(28,40)
  };                               \addlegendentry{SD-K-best (k=6)}  
 \addplot[color=green , dashed,  mark size=3pt,mark=star,line width=1pt] coordinates{
(18,135)(20,88)(22,56)(24,40)(26,30)(28,25)
  };                               \addlegendentry{SD-K-best (k=2)}  
 \addplot[color=red, dashed,  mark size=3pt,mark=star,line width=1pt] coordinates{
(18,106)(20,64)(22,39)(24,24)(26,18)(28,14)
  };                               \addlegendentry{SD-K-best (k=1)}
	 \addplot[color=blue,  mark size=3pt,mark=star,line width=1pt] coordinates{
(18,481)(20,481)(22,481)(24,481)(26,481)(28,481)
  };                               \addlegendentry{K-best (k=32)} 
    \addplot[color=black,  mark size=3pt,mark=star,line width=1pt] coordinates{
(18,151)(20,151)(22,151)(24,151)(26,151)(28,151)
  };                               \addlegendentry{K-best (k=10)}


\end{semilogyaxis}
  \end{tikzpicture} \\
\end{subfigure}
\begin{subfigure}[b]{0.4\textwidth}
  \centering
  \begin{tikzpicture}[scale=0.55,font=\small]
    \renewcommand{\axisdefaulttryminticks}{4}
    \tikzstyle{every major grid}+=[style=densely dashed]
    \tikzstyle{every axis y label}+=[yshift=-10pt]
    \tikzstyle{every axis x label}+=[yshift=5pt]
    \tikzstyle{every axis legend}+=[cells={anchor=west},fill=white,
        at={(1.01,1)}, anchor=north west, font=\small ]
\begin{semilogyaxis}[
      xmin=18, axis on top,
  title={(b) Complexity in terms og decding time (s).},
      ymin=10^-2,
      xmax=28,
      ymax=0.2,
      grid=major,
      scaled ticks=true,
    xlabel={SNR},
    ylabel={Decoding time (s) } ]
    \addplot[color=black,  mark size=3pt,mark=star,line width=1pt] coordinates{
(18,0.014)(20,0.014)(22,0.014)(24,0.0143)(26,0.0143)(28,0.0143)
  };                               \addlegendentry{K-best (k=10)}  
    \addplot[color=blue,  mark size=3pt,mark=star,line width=1pt] coordinates{
(18,0.0531)(20,0.0531)(22,0.0531)(24,0.0531)(26,0.0531)(28,0.0531)
  };                               \addlegendentry{K-best (k=32)}  
  \addplot[color=red, dashed,  mark size=3pt,mark=star,line width=1pt] coordinates{
(18,0.11)(20,0.064)(22,0.035)(24,0.021)(26,0.0138)(28,0.01)
  };                               \addlegendentry{SD-K-best (k=1)}
 \addplot[color=green , dashed,  mark size=3pt,mark=star,line width=1pt] coordinates{
(18,0.11)(20,0.065)(22,0.038)(24,0.023)(26,0.0148)(28,0.011)
  };                               \addlegendentry{SD-K-best (k=2)}  
  \addplot[color=orange, dashed,  mark size=3pt,mark=star,line width=1pt] coordinates{
(18,0.125)(20,0.075)(22,0.046)(24,0.029)(26,0.0178)(28,0.013)
  };                               \addlegendentry{SD-K-best (k=6)}  
  \addplot[color=red, dashed,  mark size=3pt,mark=star,line width=1pt] coordinates{
(18,0.13)(20,0.081)(22,0.048)(24,0.03)(26,0.018)(28,0.013)
  };                               \addlegendentry{SD-K-best (k=8)}  
 
\end{semilogyaxis}
  \end{tikzpicture} \vskip-5mm
\end{subfigure}
  \caption{Our SD-K-best latency (a), and  complexity (b)
vs. K-best approximate approach for a $16 \times 16$ 64-QAM
MIMO.}
 \label{fig:sd-k-best2}
  \end{figure}

Figures \ref{fig:sd-k-best} and \ref{fig:sd-k-best2} show  respectively the SER performance 
and complexity results for a $16 \times 16$ MIMO system using 64-QAM modulation. 
The figures compare the results of our SD-K-best algorithm with the results of K-best algorithm.  
Our SD-K-best algorithm uses twenty threads: one as a master process with a
SD instance and nineteen as workers with K-best algorithm to accelerate the search process. 
Figure \ref{fig:sd-k-best} shows the results of our proposed SD-K-best with several configurations $k \in \{1,2,6,8\}$. 
The figure also shows the results of the K-best algorithm with two configurations k=10 and k=32.
The value of $K$ refers to the number of kept nodes at each level by K-best algorithm, as well as workers in SD-K-best. 
The error rate for K-best algorithm with ten kept nodes does not
reach the acceptable error rate ($10^{-3}$) even when reaching 26 dB. 
By increasing the value of $K$ to 32, the performance of the latter is improved to meet this requirement in 22 dB.
As compared to the K-best algorithm, our proposed SD-K-best algorithm performs better in 
terms of error rate and reach an acceptable rate even with small value of kept nodes.
When increasing the number of kept nodes, the performance of our SD-K-best 
algorithm increases until reaching stagnation, i.e., 
increasing the number of kept nodes has minor  impact on the SER.   
Moreover, our SD-K-best algorithm is able to reach acceptable error rate at a round 20 dB. 
Thus four dB improvement in power consumption as compared to the K-best algorithm with $K$ equals to 32.
The performance of our proposed approach in terms of error rate is explained by 
the large number of explored combinations as compared to a conventional K-best algorithm.
Since each worker takes a subtree from the master and explores it according to the K-best algorithm,
while taking benefit from the sphere radius to explore only promising branches which reduces the complexity. 

Figure \ref{fig:sd-k-best2} (a) shows the average number of explored nodes per thread of our 
SD-K-best algorithm against the number of explored nodes 
by K-best algorithm for a  $16 \times 16$ 64-QAM MIMO system.
In the same way, Figure \ref{fig:sd-k-best2} (b) shows the time complexity of 
our SD-K-best approach against K-best algorithm for the same MIMO system.

We can see from the Figure \ref{fig:sd-k-best2} that the complexity and 
the number of visited nodes by the K-best algorithm are fixed and stable across SNR. 
This is not the case of our SD-K-best algorithm since the average number 
of explored nodes per thread decreases when increasing the SNR. 
This behavior is closely related to the sphere radius which depends essentially on the SNR, i.e., 
higher the SNR the smaller the radius, thus reducing the number of  explored nodes and the complexity.
However, the whole number of explored nodes by all workers in SD-K-best algorithm 
is much bigger as compared to the K-best algorithm which explains the improvement in error rate. 
Moreover, Figure \ref{fig:sd-k-best2} (b) shows that the complexity of the SD-K-best algorithm 
is higher than the complexity of K-best algorithm in SNR between 18 and 22 dB region. This is due 
to the fact that the SD-K-best explores more search space than K-best algorithm, 
which explains why its complexity is higher. 
Furthermore, the high complexity is also explained by the complexity of the SD algorithm 
executed by the master process.
However, this is not the case in the high SNR region (22 dB to 26 dB) where the SD-K-best algorithm
has less complexity than K-best algorithm.  This is the results of an efficient
pruning process due to the small sphere radius in this region. In other words,
the SD-K-best in the high SNR region explores less solutions, but they are of good
quality as compared to K-best explored combinations. 
The challenge to deal with massive MIMO efficiently is based on finding the
appropriate trade-off between the complexity and the performance in terms of
error rate. To find it in our case, we combined strengths of both PSD and K-best 
algorithms to ensure low complexity and good BER at the same time.  
Unlike literature works, our proposed SD-K-best approach is able to reach both real time 
complexity and good error rate from SNR equals to 28 dB.
  
In the following, we scale-up the number of antennas to
evaluate the ability of our hybrid SD-K-best algorithm to guarantee
both low complexity and good error rate performance. 
	
	\begin{figure}
	  \centering
\begin{subfigure}[b]{0.4\textwidth}
  \centering
  \begin{tikzpicture}[scale=0.55,font=\small]
    \renewcommand{\axisdefaulttryminticks}{4}
    \tikzstyle{every major grid}+=[style=densely dashed]
    \tikzstyle{every axis y label}+=[yshift=-10pt]
    \tikzstyle{every axis x label}+=[yshift=5pt]
    \tikzstyle{every axis legend}+=[cells={anchor=west},fill=white,
        at={(1.01,1)}, anchor=north west, font=\small ]
    \begin{semilogyaxis}[
      xmin=18,
			title={(a) Error Rate.},
      ymin=10^-7,
      xmax=28,
      ymax=*10^-1,
      grid=major,
      scaled ticks=true,
      xlabel={SNR},
    ylabel={Symbol Error Rate} ]
    \addplot[color=gray,  mark size=3pt,mark=star,line width=1pt]coordinates{
(18,0.025)(20,0.013)(22,0.0067)(24,0.005)(26,0.0033)(28,0.0016)};
\addlegendentry{K-best (k=10)}
 \addplot[color=red, dashed,mark size=3pt,mark=star,line width=1pt] coordinates{(18,0.03)(20,0.016)(22,0.0036)(24,0.0015)(26,0.0007)(28,0.00015)};
\addlegendentry{SD-K-best (k=1)}
 \addplot[color=green, dashed,mark size=3pt,mark=star,line width=1pt] coordinates{
(18,0.0058)(20,0.0034)(22,0.00065)(24,0.0002)(26,0.00006)(28,0.000026)};
\addlegendentry{SD-K-best (k=2)}

 \addplot[color=orange, dashed,mark size=3pt,mark=square*,line width=1pt] coordinates{
(18,0.0055)(20,0.0012)(22,0.0003)(24,0.00008)(26,0.000003)(28,0.00000066)};
\addlegendentry{SD-K-best (k=6)}
 \addplot[color=black, dashdotted,,mark size=3pt,mark=triangle*,line width=1pt] coordinates{
(18,0.0045)(20,0.0012)(22,0.0002)(24,0.000073)(26,0.0000001)(28,0.000000)};
\addlegendentry{SD-K-best (k=8)}
\end{semilogyaxis}
  \end{tikzpicture} \vskip-5mm
\end{subfigure}
\begin{subfigure}[b]{0.4\textwidth}
  \centering
\begin{tikzpicture}[scale=0.55,font=\small]
    \renewcommand{\axisdefaulttryminticks}{4}
    \tikzstyle{every major grid}+=[style=densely dashed]
    \tikzstyle{every axis y label}+=[yshift=-10pt]
    \tikzstyle{every axis x label}+=[yshift=5pt]
    \tikzstyle{every axis legend}+=[cells={anchor=west},fill=white,
        at={(1.01,1)}, anchor=north west, font=\small ]
\begin{semilogyaxis}[
      xmin=18, axis on top,
  title={(b) Complexity in terms of decoding time (s).},
      ymin=5*10^-2,
      xmax=28,
      ymax=1.2,
      grid=major,
      scaled ticks=true,
    xlabel={SNR},
    ylabel={Decoding time (s) } ]
    \addplot[color=gray,mark size=3pt,mark=star,line width=1pt] coordinates{
(18,0.294)(20,0.294)(22,0.294)(24,0.2943)(26,0.2943)(28,0.2943)}; 
\addlegendentry{K-best (k=10)} 
\addplot[color=black,dashdotted,mark size=3pt,mark=triangle*,line width=1pt]coordinates{
(18,1.51)(20,0.87)(22,0.45)(24,0.24)(26,0.128)(28,0.097)};
\addlegendentry{SD-K-best (k=8}
\addplot[color=orange,dashed,mark size=3pt,mark=square*,line width=1pt]coordinates{
(18,1.1)(20,0.61)(22,0.32)(24,0.17)(26,0.114)(28,0.095)};
\addlegendentry{SD-K-best (k=6)}
\addplot[color=green,dashed,mark size=3pt,mark=star,line width=1pt]coordinates{
(18,0.89)(20,0.5)(22,0.25)(24,0.14)(26,0.10)(28,0.085)};
\addlegendentry{SD-K-best (k=2)}
\addplot[color=red,dashed,mark size=3pt,mark=star,line width=1pt]coordinates{
(18,0.89)(20,0.47)(22,0.25)(24,0.16)(26,0.09)(28,0.08)};
\addlegendentry{SD-K-best (k=1)}
\end{semilogyaxis}
  \end{tikzpicture} \vskip-5mm
\end{subfigure}
  \caption{Error rate and complexity of our SD-K-best and  K-best algorithms for a $100 \times 100$ MIMO system using 64-QAM modulation.}
  \label{fig:sd-k-best11}
  \end{figure}
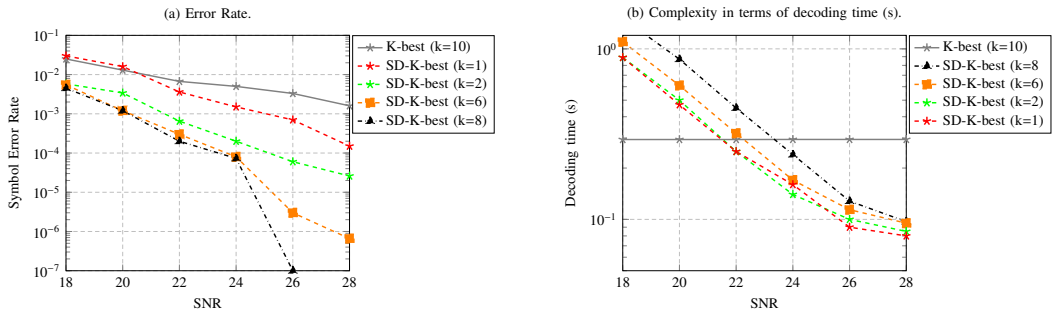

Figure \ref{fig:sd-k-best11} shows the obtained results in 
terms complexity and error rate performance, for a $100 \times 100$ MIMO system
with the 64-QAM modulation, i.e., an uncoded transmission rate of 400 bits per transmission. The first observation from the figure is the
ability of approximate algorithms to deal with large MIMO systems. 
This is not the case of the SD algorithm due to its prohibit complexity. 
Figure \ref{fig:sd-k-best11} (a) shows the large improvement in terms of error rate 
performance of our SD-k-best algorithm with low number of kept nodes, 
as compared to K-best algorithm due to the diversification gain obtained 
from using parallelization.   
Moreover, starting from 22 dB, our algorithm has a low complexity 
than k-best algorithm due to the small radius in this region, 
which induces an efficient pruning process.
Furthermore, increasing the number of kept-nodes for our SD-K-best algorithm 
have a good impact of the  error rate, without increasing much the complexity 
as shown in Figure \ref{fig:sd-k-best11} (b). 
Indeed, the diversification gain improves more efficiently the radius which  
helps to reduce the overhead of increasing the number of kept-nodes.   

In addition, the high SNR region in Figure \ref{fig:sd-k-best11} (b) shows that
the curves begin to stabilize (floor) before even reaching the 10 ms threshold. 
In fact, when dealing with a larger number of antennas and constellation sizes, 
reaching a first leaf node can take more than 10 ms.
This is due to the heavy computation needed at each level of the tree
and the limited number of computing elements in CPU architecture. 
For this reason, it is critical to use highly efficient computing architectures with
a large number of computing elements such as GPUs, which will be the subject of our future work.

\section{Conclusion}
\label{sec:summary}
We proposed in this paper the optimization of a well
known optimal decoder with high complexity named, the Sphere Decoder
(SD) algorithm. To achieve this goal, three levels have been 
considered. The first level consisted in optimizing the SD 
complements, especially the exploration strategies and the evaluation
process since they have a huge impact on the complexity.
Since the search tree for all the possible combination of the
transmitted vector is huge, the second level aimed to speedup the
search-tree exploration process by using parallel architectures.  
Finally, the third level consisted to use approximate algorithms
that perform a trade-off between the complexity and the performance.
The obtained results in each level confirmed our proposals and allowed
us, not only to speedup the SD algorithm by a factor of $60\times$ using a 16-QAM modulation, but also
to deal with large MIMO systems with dense constellations, such as $100\times100$.
To conclude this work, the challenge to deal with massive 
MIMO efficiently is based on finding the
appropriate trade-off between the complexity and the performance in terms of
error rate. To find it in our case, we combined the strengths of both parallel SD and K-best 
algorithms to ensure a low complexity and good error rate performance at the same time.

In the future, we plan to explore more parallel approximate algorithms
to deal with a multi-user case situation using GPU architectures.\\


\begin{spacing}{1}
\bibliographystyle{acm} 
\bibliography{references}

\begin{thebibliography}{10}

\bibitem{agrell2002closest}
{\sc Agrell, E., Eriksson, T., Vardy, A., and Zeger, K.}
\newblock Closest point search in lattices.
\newblock {\em IEEE transactions on information theory 48}, 8 (2002),
  2201--2214.

\bibitem{arfaoui2016}
{\sc Arfaoui, M.-A., Ltaief, H., Rezki, Z., Alouini, M.-S., and Keyes, D.}
\newblock {Efficient Sphere Detector Algorithm for Massive MIMO Using GPU
  Hardware Accelerator}.
\newblock {\em Procedia Computer Science 80\/} (2016), 2169 -- 2180.
\newblock International Conference on Computational Science 2016, ICCS 2016,
  6-8 June 2016, San Diego, California, USA.

\bibitem{bjornson2016massive}
{\sc Bj{\"o}rnson, E., Larsson, E.~G., and Marzetta, T.~L.}
\newblock Massive mimo: Ten myths and one critical question.
\newblock {\em IEEE Communications Magazine 54}, 2 (2016), 114--123.

\bibitem{blas}
{\sc BLAS.}
\newblock Basic linear algebra subprograms. http://www.netlib.org/blas, 2013.

\bibitem{chen2015gpu}
{\sc Chen, T., and Leib, H.}
\newblock Gpu acceleration for fixed complexity sphere decoder in large mimo
  uplink systems.
\newblock In {\em IEEE 28th Canadian Conference on Electrical and Computer
  Engineering (CCECE), 2015\/} (2015), IEEE, pp.~771--777.

\bibitem{dahlman20103g}
{\sc Dahlman, E., Parkvall, S., Skold, J., and Beming, P.}
\newblock {\em 3G evolution: HSPA and LTE for mobile broadband}.
\newblock Academic press, 2010.

\bibitem{fincke1985improved}
{\sc Fincke, U., and Pohst, M.}
\newblock Improved methods for calculating vectors of short length in a
  lattice, including a complexity analysis.
\newblock {\em Mathematics of computation 44}, 170 (1985), 463--471.

\bibitem{foschini1996layered}
{\sc Foschini, G.~J.}
\newblock Layered space-time architecture for wireless communication in a
  fading environment when using multi-element antennas.
\newblock {\em Bell labs technical journal 1}, 2 (1996), 41--59.

\bibitem{hassibi2005sphere}
{\sc Hassibi, B., and Vikalo, H.}
\newblock On the sphere-decoding algorithm i. expected complexity.
\newblock {\em IEEE transactions on signal processing 53}, 8 (2005),
  2806--2818.

\bibitem{husmann2017flexcore}
{\sc Husmann, C., Georgis, G., Nikitopoulos, K., and Jamieson, K.}
\newblock Flexcore: Massively parallel and flexible processing for large
  $\{$MIMO$\}$ access points.
\newblock In {\em 14th $\{$USENIX$\}$ Symposium on Networked Systems Design and
  Implementation ($\{$NSDI$\}$ 17)\/} (2017), pp.~197--211.

\bibitem{jozsa2013new}
{\sc J{\'o}zsa, C.~M., Kolumb{\'a}n, G., Vidal, A.~M.,
  Mart{\'\i}nez-Zald{\'\i}var, F.-J., and Gonz{\'a}lez, A.}
\newblock New parallel sphere detector algorithm providing high-throughput for
  optimal mimo detection.
\newblock {\em Procedia Computer Science 18\/} (2013), 2432--2435.

\bibitem{mimo-iot}
{\sc Lee, B.~M., and Yang, H.}
\newblock Massive mimo for industrial internet of things in cyber-physical
  systems.
\newblock {\em IEEE Transactions on Industrial Informatics 14}, 6 (June 2018),
  2641--2652.

\bibitem{lu2014overview}
{\sc Lu, L., Li, G.~Y., Swindlehurst, A.~L., Ashikhmin, A., and Zhang, R.}
\newblock An overview of massive mimo: Benefits and challenges.
\newblock {\em IEEE journal of selected topics in signal processing 8}, 5
  (2014), 742--758.

\bibitem{marzetta2015massive}
{\sc Marzetta, T.~L.}
\newblock Massive mimo: an introduction.
\newblock {\em Bell Labs Technical Journal 20\/} (2015), 11--22.

\bibitem{mccann2014official}
{\sc McCann, S., and Ashley, A.}
\newblock Official ieee 802.11 working group project timelines.
\newblock {\em November2011.[Online]. Available:
  http://www.ieee802.org/11/Reports/802.11\_Timelines.htm\/} (2014).

\bibitem{nikitopoulos2018massively}
{\sc Nikitopoulos, K., Georgis, G., Jayawardena, C., Chatzipanagiotis, D., and
  Tafazolli, R.}
\newblock Massively parallel tree search for high-dimensional sphere decoders.
\newblock {\em IEEE Transactions on Parallel and Distributed Systems\/} (2018).

\bibitem{nikitopoulos2014geosphere}
{\sc Nikitopoulos, K., Zhou, J., Congdon, B., and Jamieson, K.}
\newblock Geosphere: Consistently turning mimo capacity into throughput.
\newblock {\em ACM SIGCOMM Computer Communication Review 44}, 4 (2014),
  631--642.

\bibitem{paulraj1994increasing}
{\sc Paulraj, A.~J., and Kailath, T.}
\newblock Increasing capacity in wireless broadcast systems using distributed
  transmission/directional reception (dtdr), Sept.~6 1994.
\newblock US Patent 5,345,599.

\bibitem{raleigh1998spatio}
{\sc Raleigh, G.~G., and Cioffi, J.~M.}
\newblock Spatio-temporal coding for wireless communication.
\newblock {\em IEEE Transactions on communications 46}, 3 (1998), 357--366.

\bibitem{roger2012fully}
{\sc Roger, S., Ramiro, C., Gonzalez, A., Almenar, V., and Vidal, A.~M.}
\newblock Fully parallel gpu implementation of a fixed-complexity soft-output
  mimo detector.
\newblock {\em IEEE Transactions on Vehicular Technology 61}, 8 (2012),
  3796--3800.

\bibitem{Marvin-book}
{\sc Simon, M.~K., and Alouini, M.-S.}
\newblock {\em Digital Communication over Fading Channels (Wiley Series in
  Telecommunications and Signal Processing)}, 2nd~ed.
\newblock {Wiley-IEEE Press}, New York, NY, USA, 2004.

\bibitem{su2013investigation}
{\sc Su, X., Zeng, J., Rong, L.-P., and Kuang, Y.-J.}
\newblock Investigation on key technologies in large-scale mimo.
\newblock {\em Journal of Computer Science and Technology 28}, 3 (2013),
  412--419.

\bibitem{telatar1999capacity}
{\sc Telatar, E.}
\newblock Capacity of multi-antenna gaussian channels.
\newblock {\em European transactions on telecommunications 10}, 6 (1999),
  585--595.

\bibitem{viterbo1999universal}
{\sc Viterbo, E., and Boutros, J.}
\newblock A universal lattice code decoder for fading channels.
\newblock {\em IEEE Transactions on Information theory 45}, 5 (1999),
  1639--1642.

\bibitem{wong2002vlsi}
{\sc Wong, K.-w., Tsui, C.-y., Cheng, R.-K., and Mow, W.-h.}
\newblock A vlsi architecture of a k-best lattice decoding algorithm for mimo
  channels.
\newblock In {\em Circuits and Systems, 2002. ISCAS 2002. IEEE International
  Symposium on\/} (2002), vol.~3, IEEE, pp.~III--III.

\bibitem{wu2012flexible}
{\sc Wu, M., Yin, B., and Cavallaro, J.~R.}
\newblock Flexible n-way mimo detector on gpu.
\newblock In {\em 2012 IEEE Workshop on Signal Processing Systems\/} (2012),
  IEEE, pp.~318--323.

\bibitem{wu2014gpu}
{\sc Wu, M., Yin, B., Wang, G., Studer, C., and Cavallaro, J.~R.}
\newblock Gpu acceleration of a configurable n-way mimo detector for wireless
  systems.
\newblock {\em Journal of Signal Processing Systems 76}, 2 (2014), 95--108.

\bibitem{zheng2015survey}
{\sc Zheng, K., Zhao, L., Mei, J., Shao, B., Xiang, W., and Hanzo, L.}
\newblock Survey of large-scale mimo systems.
\newblock {\em IEEE Communications Surveys \& Tutorials 17}, 3 (2015),
  1738--1760.

\end{thebibliography}
\end{spacing}


\end{document}